\documentclass[letter,fleqn]{cas-sc}

\usepackage[authoryear,longnamesfirst]{natbib}

\def\tsc#1{\csdef{#1}{\textsc{\lowercase{#1}}\xspace}}
\tsc{WGM}
\tsc{QE}

\usepackage{amssymb}
\usepackage{amsbsy}
\usepackage{amsmath}
\usepackage{array}
\usepackage{color}
\usepackage{psfrag,epsfig}
\usepackage{subfig}
\usepackage{enumitem}
\usepackage{nicefrac, xfrac}
\usepackage{multirow}
\definecolor{mygreen}{RGB}{0,128,0}
\newcommand{\vx}{{\mathbf x}}

\newcommand{\be}{\begin{equation}}
\newcommand{\ee}{\end{equation}}

\newcommand{\pe}{P\'eclet }
\newcommand{\da}{Damk\"ohler }

\begin{document}
\let\WriteBookmarks\relax
\def\floatpagepagefraction{1}
\def\textpagefraction{.001}

\shorttitle{Learning the Factors Controlling Mineralization for GCS} 

\shortauthors{A. Pachalieva, J. D. Hyman, D. O'Malley, H. Viswanathan, G. Srinivasan}

\title[mode = title]{Learning the Factors Controlling Mineralization for Geologic Carbon Sequestration}  



%

%
\author[1,2]{Aleksandra Pachalieva}

\cormark[1]


\ead{apachalieva@lanl.gov}



\author[2]{Jeffrey D. Hyman}
\ead{jhyman@lanl.gov}

\author[2]{Daniel O'Malley}
\ead{omalled@lanl.gov}

\author[2]{Hari Viswanathan}
\ead{viswana@lanl.gov}

\author[3]{Gowri Srinivasan}
\ead{gowri@lanl.gov}

\affiliation[1]{organization={Center for Nonlinear Studies (CNLS),Theoretical Division, Los Alamos National Laboratory},
            city={Los Alamos},
            postcode={87544}, 
            state={NM},
            country={USA}}
            
\affiliation[2]{organization={Energy \& Natural Resources Security Group (EES-16), Earth and Environmental Sciences Division, Los Alamos National Laboratory},
            city={Los Alamos},
            postcode={87544}, 
            state={NM},
            country={USA}}

\affiliation[2]{organization={X Computational Physics Division, Los Alamos National Laboratory},
            city={Los Alamos},
            postcode={87544}, 
            state={NM},
            country={USA}}

\cortext[1]{Corresponding author}



\begin{abstract}
We perform a set of flow and reactive transport simulations within three-dimensional fracture networks to learn the factors controlling mineral reactions. CO$_2$ mineralization requires CO$_2$-laden water, dissolution of a mineral that then leads to precipitation of a CO$_2$-bearing mineral. Our discrete fracture networks (DFN) are partially filled with quartz that gradually dissolves until it reaches a quasi-steady state. At the end of the simulation, we measure the quartz remaining in each fracture within the domain. We observe that a small backbone of fracture exists, where the quartz is fully dissolved which leads to increased flow and transport. However, depending on the DFN topology and the rate of dissolution, we observe a large variability of these changes, which indicates an interplay between the fracture network structure and the impact of geochemical dissolution. In this work, we developed a machine learning framework to extract the important features that support mineralization in the form of dissolution. In addition, we use structural and topological features of the fracture network to predict the remaining quartz volume in quasi-steady state conditions. As a first step to characterizing carbon mineralization, we study dissolution with this framework. We studied a variety of reaction and fracture parameters and their impact on the dissolution of quartz in fracture networks. We found that the dissolution reaction rate constant of quartz and the distance to the flowing backbone in the fracture network are the two most important features that control the amount of quartz left in the system. For the first time, we use a combination of a finite-volume reservoir model and graph-based approach to study reactive transport in a complex fracture network to determine the key features that control dissolution.  
\end{abstract}




\maketitle


\section{Introduction}

There is increasingly clear evidence of climate change and a strong desire in the research community to combat the effects through techniques such as reducing emissions, carbon dioxide (CO$_2$) sequestration, as well as using renewable energy and conserving resources. One of the promising climate change mitigation strategies is CO$_2$ sequestration, where CO$_2$ is captured from the atmosphere, and then injected and stored in deep saline aquifers. Injecting large amounts of CO$_2$ into geologic formations can cause a range of coupled processes such as thermal, mechanical, hydrodynamic, and chemical \cite{white2003separation,gaus2010role,shao2010dissolution}. This brings the minerals and the waters in the rock formation out of equilibrium, which may cause the dissolution of rock minerals and/or precipitation of secondary minerals and thus change the properties of the pre-existing rock such as morphology, porosity, and permeability. These processes affect the flow and transport of CO$_2$ and water within the reservoir and are critical for the integrity of the long-term fate of the geologic carbon sequestration (GCS). Therefore, it is critical to understand the geophysical and chemical interactions between minerals and fluids, in order to ensure the efficiency and sustainability of the GCS. 

In fairly homogeneous rock formations, reactive fronts can be adequately modeled using reaction rates constrained by elementary principles~\cite{Maher2009SecMinerals,Moore2012RTM,Navarre-Sitchler2011ABasalt,White2008ChemicalProfiles}.
However, within fracture networks the application of these simple models is limited, since there exists a highly heterogeneous fluid flow field due to the spatially variable resistance offered to flow by the geo-structural attributes of the fracture network~\cite{hyman2018dispersion,hyman2019emergence,hyman2020characterizing,maillot2016connectivity,kang2020anomalous,painter2002power,neuman2005trends,sherman2018characterizing,sweeney2020stress,sweeney2023characterizing,yoon2023effects}.
As readily accessible reactive minerals are depleted, the apparent, or domain-averaged, mineral dissolution rate decreases to values that can be orders of magnitude lower than the laboratory-measured rate~\cite{andrews2021temporal,atchley2013using,beisman2015PCF,jung2018physical}.
In turn, reactions in fractured media are often transport-controlled and elementary models cannot properly constrain/predict reaction rates~\cite{andrews2021temporal,andrews2023fracture,berkowitz1997anomalous,becker2000tracer,edery2016structural,geiger2010upscaling,haggerty2001tracer,huseby2001dispersion,hyman2019matrix,jung2018physical,jung2018scale,pandey2016modeling,kang2020anomalous,Meigs2001,painter2002power,WEN20181}. 
Characterizing the feedback between the network structure on the flow field and associated reactive transport requires a coupled thermo-hydro-chemical simulator capable of dynamically modifying flow resistance (hydraulic aperture/permeability) within a three-dimensional fracture network. 
To date, most computational studies of geochemical reactions have been carried out in a single fracture, small two-dimensional networks, or in upscaled/equivalent continuum models \cite{andrews2021temporal,andrews2023fracture,Deng2018Pore-scaleFractures,feng2019mineral,Lebedeva2017WeatheringSystems,jones2019mineral,Molins2019Multi-scaleRates,noiriel2021geometry,pandey2016modeling,Steefel1998MulticomponentGeometry,steefel1994coupled,steefel2022reactive}.
These three-dimensional high-fidelity simulations, although heavily sought after, were relatively infeasible due to computational limitations. 
However, recent developments in high-performance computing now allow for the exploration of flow and reactive transport properties in 3D fractured media~\cite{hyman2022geo}.


Machine learning (ML) techniques have shown tremendous promise in geosciences due to the ability to infer parameters and mechanisms of importance with relatively low computational burden. Previous research has exploited ML algorithms to construct reduced order models (ROM) which once trained, can run in a fraction of the time it takes to solve complex advection-dispersion-reaction (ADR) equations \cite{santiago2014methodology, goetz2015evaluating, valera2017machine}. However, this is only possible because problems including characterizing CO$_2$ mineralization lend themselves very easily to being represented on a lower dimensional manifold. Flow and transport through fractured rock often happen in preferential pathways where the majority of the reservoir does not participate in the first-order physics. The backbone of the fracture network where the majority of reactive flow takes place is a small fraction of the entire domain of interest. Inferring the characteristics of the backbone prior to solving the ADR yields significant savings in computational time \cite{srinivasan2019model,viswanathan2018advancing}\cite{vesselinov2019unsupervised,ahmmed2021comparative,liu2022machine}. Structured systems such as fractured rock formations have also been successfully represented as pipes or graphs in previous research, which allows for solving flow and transport equations on the network rather than relying on expensive meshing constructs. 

One other benefit of the surrogate models or ROMs obtained through training ML algorithms is their use in a multi-fidelity uncertainty quantification sense. Several thousands to millions of lower fidelity ML algorithms can be used in conjunction with a handful of high fidelity runs to improve both precision and accuracy of the predictions \cite{omalley2018efficient}. 

It is challenging to determine the interplay between the network geostructure and geochemical reactions in fractured media. Even though our physics-based models can simulate these coupled processes, it is difficult to unravel which are the model parameters that control key quantities of interest (e.g. amount of quartz remaining or CO$_2$ mineralized in a given fracture).
In this work, we perform a large number of flow and reactive transport simulations in three-dimensional discrete fracture networks (DFN) to determine the important features that control geochemical mineralization in the form of dissolution. We isolate the effects of mineral dissolution, by focusing on the simple reaction of quartz dissolution taking into account that this reaction does not induce significant pH changes and will not be sensitive to pH changes in the fluid as other mineral reactions might be. We construct a set of fracture networks composed of a single families of mono-disperse disc-shaped fractures. Initially, the fractures are partially filled with quartz, which dissolves gradually until a quasi-steady state is reached. Depending on the DFN topology and the rate of dissolution, we observe large discrepancies in the remaining quartz in each fracture at the end of the reactive transport simulation. This indicates an interplay between the fracture network structure and the impact of geochemical dissolution, which we would like to understand better. 
Determining where dissolution or precipitation occurs in a network is critical since they could be positive or negative feedbacks to the flow. Some fractures may flow better allowing for more access to the formation in the case of dissolution. For precipitation, fractures may clog blocking parts of the reservoir. To achieve this, we combine graph representation of the DFNs and machine learning techniques, that allow us to predict the remaining quartz volume in quasi-steady state conditions using the following three categories of features of the fracture network: topological, geometric, and hydrological. By using a regression model, we are also able to assess the importance of different features that characterize the fracture networks. We observe that the most important features are the topological and geometric features, and they are sufficient to train a regression model that predicts the remaining quartz volume in the system. We chose the remaining quartz volume as quantity of interest in our study, because it tells us where most of the reaction is taking place. The distance to the backbone and the rate constant are the features which significantly affect the amount of quartz remaining in each fracture, which is understandable since they control the flow channelization and the strength of the chemical reaction, respectively. The hydrological quantities are the least important ones with respect to the amount of quartz remaining in the system; however, including them in the training process leads to improved confidence in the regression model. 

For future work, we aim to simulate the effects of precipitation and dissolution together to be able to fully characterize carbon mineralization. However, looking at dissolution is the first step in testing out this framework.

In Section \ref{sec:methods}, we describe the methods used to generate the DFNs and simulate flow and reactive transport, as well as the machine learning methods we use. Additionally, we introduce the important features (geometrical, topological, and hydrological) that we use to characterize the fracture networks. In Section \ref{sec:results}, we describe the main results of our work, including the random forest model training and testing results, as well as the results from an extensive grid search. In Section \ref{sec:discussion}, we discuss the implications of our results and provide conclusions.

\section{Methods: Computational Approach}\label{sec:methods}

Our primary goal is understanding the connection between properties of the fracture network (topological, geometric, and hydraulic) and resulting geochemical reactions within the network. 
Identifying such connections is difficult because of the complexity of both the fracture network structure and it's impact on fluid flow properties, which determines transport and subsequent reactions. 
To do so, we adopt a multi-fidelity computational approach. 
We consider three-dimensional fracture networks using a discrete fracture network (DFN) method.
Discrete fracture network models are distinguished from continuum models in that the fractures are explicitly represented rather than via their upscaled effective properties~\cite{davy2013model,davy2010likely,de2004influence,dreuzy2012influence,erhel2009flow,hyman2014conforming,hyman2015dfnworks,Flemisch2016,manzoor2018interior,hyman2022flow}.
Due to the disparity between their length and aperture, DFN models represent fractures as co-dimension one objects, e.g., lines in two-dimensional simulations and planes in three-dimensional simulations. 
Each fracture is assigned a shape, location, orientation, and hydraulic properties based on field site characterization. 
The individual fractures interconnect to form a network.
We characterize these fracture networks in terms of their topological (connectivity), geometrical, and hydrological properties using a graph-based approach~\cite{hyman2018identifying}. 
Graph-based approaches have been shown to be a critical companion to DFN simulations are they provide a rigorous and interpretable manner in which to link fracture network properties to flow and reactive transport observations~\cite{hyman2017predictions,hyman2018dispersion,hyman2019linking,hyman2020flow,hyman2021scale,pachalieva2023impact,srinivasan2018quantifying,yoon2023effects}.
Flow and reactive transport within the networks are simulated using the massively parallel subsurface flow and reactive transport code {\sc pflotran}~\cite{lichtner2015pflotran}.
There are a variety of reactive transport simulators available that differ in terms of spatial dimensions, discretization schemes, time integration methods, governing equations, flow simulator capabilities (single-phase Darcy flow, variable saturation Richards flow, multi-phase flow, variable density, non-isothermal, and heterogeneous permeability), transport formulations (advection, mechanical dispersion, molecular diffusion, multi-continuum), and geochemistry options (surface complexation, kinetic mineral precipitation-dissolution, aqueous kinetics, mineral nucleation, mineral solid-solutions).
The most common simulators in use today are {\sc PHREEQC}~\cite{parkhurst2013description} (which is the geochemistry engine for {\sc HPx}~\cite{jacques2018hpx}, {\sc PHT3D}~\cite{prommer2003modflow}, and {\sc OpenGeoSys}~\cite{kolditz2012opengeosys}), HYTEC~\cite{van2003module}, {\sc ORCHESTRA}~\cite{meeussen2003orchestra}, {\sc TOUGHREACT}~\cite{xu2011toughreact}, {\sc eSTOMP}~\cite{white2003stomp}, {\sc HYDROGEOCHEM}~\cite{yeh1990hydrogeochem}, {\sc CrunchFlow}~\cite{steefel2009crunchflow}, {\sc MIN3P}~\cite{su2021min3p}, and {\sc PFLOTRAN}~\cite{lichtner2015pflotran}. 
A comparison of strengths and weaknesses between the codes is provided in~\citet{steefel2015reactive}. 
The aforementioned graph-based approaches for network characterization have been combined with machine learning techniques to link geo-structure with flow and transport~\cite{valera2018machine,srinivasan2019model,srinivasan2020physics}, but this study marks the first time to do so in the context of reactive transport models. 

In the first portion of this section, we describe the geometric simulations for reactive transport modeling in fractured media which are used to generate our data set. 
Next, we describe our adopted machine learning techniques.

\subsection{Reactive transport modeling}

Our reactive transport modeling through fractured media is broken into two steps. 
The first step is the generation of an ensemble of generic three-dimensional DFN models. 
In the DFN methodology, fractures are explicitly represented as planar polygons in space. 
Each fracture has a stochastically sampled shape, size, and orientation.
The fracture interconnects to form a network through which flow occurs.
This is in contrast to continuum models, where fractures are represented by their effective properties\cite{sweeney2020upscaled}. 
The choice to use a DFN model, rather than a continuum model, is born from the desire to link reactive transport observations directly with fracture attributes.
The second step is simulating the flow field and associated reactive transport, which requires a coupled thermo-hydro-chemical simulator capable of dynamically modifying flow resistance (hydraulic aperture/permeability) within a three-dimensional fracture network. 

\subsubsection{Three-Dimensional Discrete Fracture Network Modeling}
We use {\sc dfnWorks}~\cite{hyman2015dfnworks} software suite to perform our reactive transport simulations in three-dimensional discrete fracture networks (DFN), cf.~\cite{viswanathan2022from} for a comprehensive discussion of DFN modeling approaches.
We consider a set of generic networks composed of a single families of mono-disperse (constant sized) disc-shaped fractures in a cubic domain with sides of length 10 meters. 
Each fracture has a radius of 1.5 meters and their centers are uniformly distributed throughout the domain.
Fractures are placed into the domain until a fracture intensity, total surface area over total volume, of $P_{32}$ = 3.25 is obtained. 
During generation, the domain is increased by 0.5 meters in all directions to mitigate boundary density effects. 
The orientation of the fracture family is randomly distributed across the unit sphere to mimic a disordered media~\cite{hyman2018dispersion}. 
The generic nature of these parameters is designed to isolate the effects of reactive transport and network connectivity from other structural attributes of fracture networks. 
After generation is complete, we remove isolated fractures, because they do not participate in flow.
Once the final network is produced, the feature generation for meshing ({\sc fram}) described in~\cite{hyman2014conforming}, which combines network generation with mesh generation to remove features that degrade mesh quality, and the near-Maximal Algorithm for Poisson Sampling (nMAPS) presented in~\cite{krotz2022maximal}, are implemented using the LaGriT meshing toolbox~\cite{lagrit2011} to generate a conforming Delaunay triangulation, i.e.,  the computational mesh. 
The mesh is composed of uniform triangular elements with edge lengths of 0.05 m. 
Initially, all fractures are assigned a uniform  hydraulic aperture of $b = 1 \cdot 10^{-5}$ m, which is a reasonable value in crystalline rocks such as granite for fractures of this size~\cite{SKB2010}.
The cubic law is used to define the initial permeability of each control volume within the fractures, $k = 8.3\cdot 10^{-12}$ m$^2$.
Even though the apertures are initially uniform, they can vary spatially.
Additional information about the inclusion of in-fracture aperture variability into DFN meshes using {\sc dfnworks} can be found in~\cite{karra2015effect,frampton2019advective,makedonska2016evaluating,hyman2021scale}.

\subsubsection{Flow and reactive transport}
\label{subsec:flow_react_transport}
Flow and reactive transport within the networks are simulated using the massively parallel subsurface flow and reactive transport code {\sc pflotran}~\cite{lichtner2015pflotran}.
Flow in the fracture network is modeled using the Richards equation with a spatially variable permeability field.
We use {\sc pflotran}~\cite{lichtner2015pflotran} to numerically integrate the governing equations for pressure and volumetric flow rates.
We use a direct method to obtain the solution to the linear system for improved accuracy and performance~\cite{greer2022comparison}.

The DFNs are initialized with an 80\% volume fraction of quartz (SiO$_2$) e.g., a uniform fracture porosity of 0.2.
Primarily quartz-filled fractures are plausible in crystalline rocks~\cite{fisher1992models,navarre2015porosity}.
A volumetric flow rate boundary condition is applied to the inflow and outflow boundaries of the domain. 
The flow rate is equal to the initial volume of the particular fracture network per year.
Even though the volume of the network increases due to dissolution, this flow rate is held constant.
As fresh water is introduced into the domain, the quartz dissolves to produce aqueous silica (SiO$_{2(aq)}$)
\begin{equation}\label{eq:quartz_diss}
SiO_{2(qz)} \rightarrow SiO_{2(aq)} 
\end{equation}
The quartz specific surface area ($A$) was defined as 0.0225 m$^2$ g$^{-1}$ ~\cite{Wollast1988} to calculate the initial total quartz surface area (A$_o$) in each cell.
Initial fluid composition in the fractures was in equilibrium with quartz and contained 1$\cdot10^{-20}$ M non-reactive tracer while the input fluid composition flowing into the domain consisted of 1$\cdot10^{-3}$ M non-reactive tracer and 1$\cdot10^{-20}$ M SiO$_{2(aq)}$.
The non-reactive tracer is used to initialize the transport equations and is not analyzed or considered in this study.
We consider four rate constants ($k$) of 1$\cdot10^{-9}$, 1$\cdot10^{-10}$, 1$\cdot10^{-11}$, 1$\cdot10^{-12}$ mol m$^{-2}$ s$^{-1}$. 

The local dissolution rate of quartz ($R$, mol m$^{-3}$ s$^{-1}$) at each time step was calculated according to linear transition state theory (TST), 
\begin{equation}\label{eq:rate}
    R=kA\left(1-\frac{SiO_{2(aq)}}{K_{eq}}\right)\,,
\end{equation}
where K$_{eq}$=10$^{-3.9993}$ at 25$^o$C, as defined in the \texttt{hanford.dat} database distributed with {\sc pflotran}. 
As quartz dissolves throughout the simulation, the volume of quartz in each cell is updated and the bulk quartz surface area in each cell is recalculated where $A$ at time \textit{t} is scaled by the ratio of quartz volume ($V$) at time \textit{t} and initial quartz volume (\textit{V$_o$})
\begin{equation}\label{eq:area}
    A_{t}=\left(\frac{Vt}{V_{o}}\right)^{2/3}A_{o}\,.
\end{equation}
The permeability, porosity, and mineral surface area of every cell in the mesh are also updated at every time step.
Permeability and mineral surface area are updated due to mineral dissolution reactions through the change in porosity
\begin{equation}
\varphi = 1-\sum_m\varphi_m.
\end{equation}
Change in permeability involves a phenomenological relation with porosity
\begin{equation}
k = k_0 f(\varphi,\,\varphi_0,\,\varphi_c,\,a),
\end{equation}
where  $k_0 $ is the initial permeability and
\begin{equation}
f =
    \begin{cases}
         \left(\frac{\varphi-\varphi_c}{\varphi_0-\varphi_c}\right)^a & \varphi > \varphi_c,\\
        f_{\rm min} & \varphi \leq \varphi_c
    \end{cases}
\end{equation}
The mineral surface area  evolves according to
\begin{equation}
A_m = A_m^0 \left(\frac{\varphi_m}{\varphi_m^0}\right)^n  \left(\frac{1-\varphi}{1-\varphi_0}\right)^{n'},
\end{equation}
where the super/subscript $0$ denotes initial values, with a typical value for $n$ of 2/3 reflecting the surface to volume ratio. Note that this relation only applies to primary minerals ($\varphi_m >0$). The quantity $\varphi_c$ refers to a critical porosity below which the permeability is assumed to be constant with scale factor $f_{\rm min}$.
Note that because hydraulic aperture is integrated into the cell volume which does not change, it is not directly updated at every step.
However, these changes in porosity and permeability would correspond to an implicit change in hydraulic aperture with constant porosity under the assumption of a dependency of permeability on the hydraulic aperture. 
Flow simulations are run for 10 million years, whence all simulations have reached a quasi-steady state in terms of the outflowing $SiO_{2(aq)} $ values. 

\begin{figure}[t]
    \centering
    \includegraphics[width=0.65\textwidth]{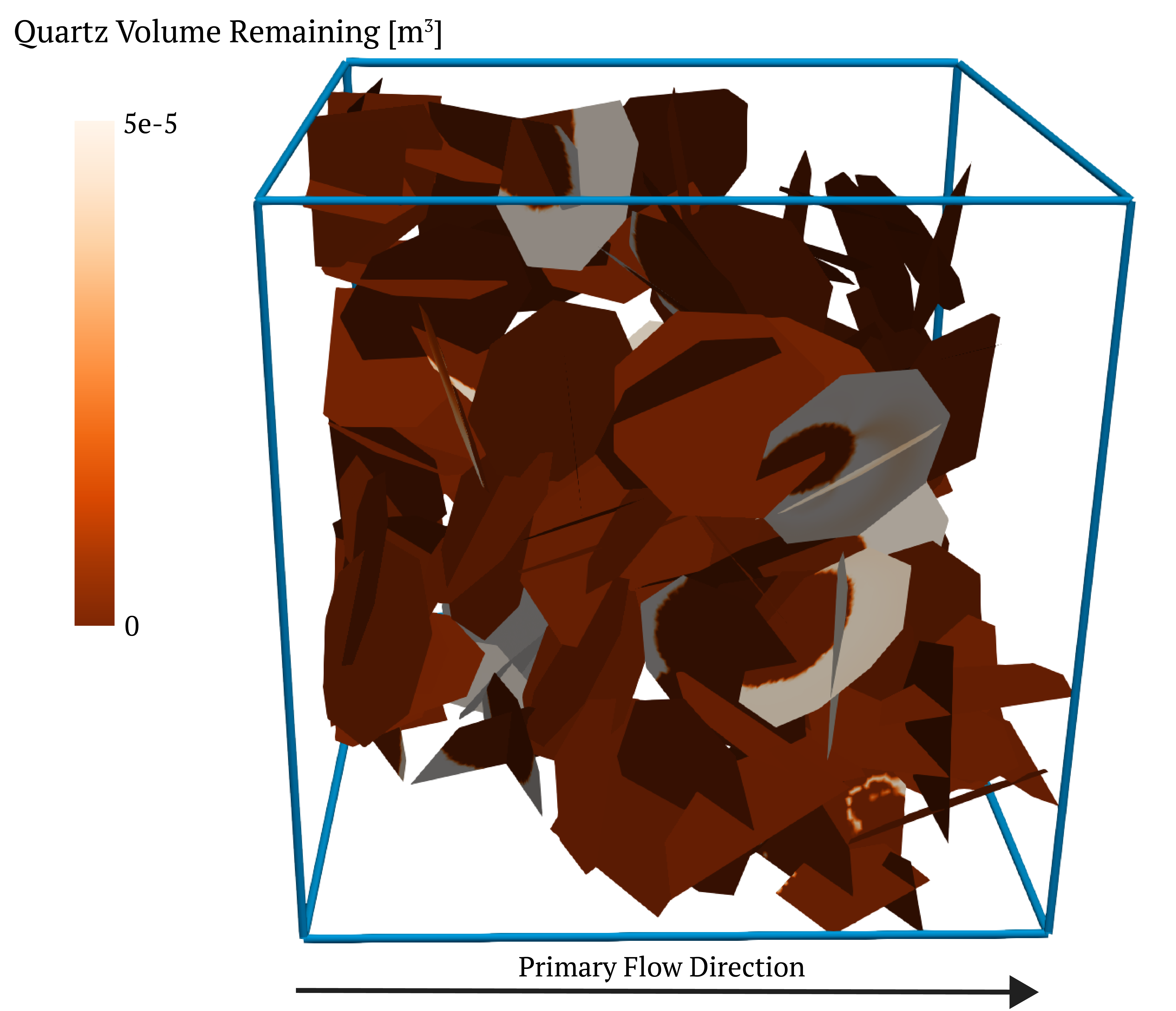}\\
    \caption{\label{fig:dfn} DFN containing 59 fractures. Fractures are colored by the remaining total quartz volume in the network at the end of the simulation.}
\end{figure}

At the end of the simulation, we measure the quartz remaining in each fracture within the domain. 
We consider both the total quartz volume [m$^3$] and the quartz volume fraction, which is the total quartz volume divided by the fracture volume. 
Figure~\ref{fig:dfn} shows one DFN from our ensemble. Fractures are colored by the total quartz volume remaining at the end of the simulation.

\subsection{Machine Learning Methods}

In this study, our data set is built of the fractures from 800 networks, each executed using one of the four rate constants as defined in Section \ref{subsec:flow_react_transport}. The data set is divided into a training set and a test set. We are using a random forest regression model implemented as part of the \texttt{scikit-learn} machine learning package in Python as a \texttt{RandomForestRegressor} function. 

The random forest method is an ensemble learning method that can be used for classification, regression, and other tasks by constructing an ensemble of recursively defined decision trees at training time. Each tree uses a random portion of the original data as training and in the case of random forest regression tasks, the output of the method is the mean or average prediction of the individual trees \cite{watt2020machine}. Random decision forests were first introduced by \citet{ho1995random}, who found that by combining forests of trees can produce a single high-performing model. 

Furthermore, random forest algorithms also provide an estimate of the importance of each individual feature, also called \textit{permutation importance}. To measure the feature's importance the values of each feature are permuted, generating new trees, and the \textit{out-of-bag error} is calculated on the new perturbed data set. The \textit{out-of-bag error} is used to validate random forest methods by averaging the error using predictions from samples that are not used in the training of a particular tree (\textit{out-of-bag samples}). The importance score is calculated by averaging the difference between the out-of-bag error before and after the permutation over all trees. If a feature is important for the regression, the permutations will produce many errors, while if a feature is not important, the permutations will not have a large effect on the performance of the regression method. 

We use the random forest regression model implementation from Python's \texttt{scikit-learn} package, in particular the \texttt{RandomForestRegressor} function. The \texttt{RandomForestRegressor} function has a number of input parameters including the following:

\begin{itemize}
    \item \textit{max\_depth}: maximum depth of the decision tree. The default value is \texttt{None}, which means that the tree is expanded until all leaves are pure or until all leaves contain less than the set minimum number of samples required to be on a leaf node.
    \item \textit{max\_features}: maximum number of features that the model will consider when determining a split. The default value is \texttt{1.0} which translates to the number of features to be considered equal to the number of features seen during the training.
    \item \textit{n\_estimators}: number of decision trees used in the training. The suggested default value is \texttt{100}.
    \item \textit{min\_samples\_leaf}: minimum number of samples required to be at a leaf node. The suggested default value is \texttt{2}.
    \item \textit{min\_samples\_split}: minimum number of samples required to split an internal node. The suggested default value is \texttt{1}.
\end{itemize}
More information on other input parameters can be found in the \texttt{scikit-learn}'s \texttt{RandomForestRegressor} documentation \cite{sklearn_rf_doc}.

\subsubsection{Performance Measures}
To assess the performance of the random forest regression model, we use two statistical measures: (1) the \textit{R-squared score} (R2), also called \textit{coefficient of determination}; and (2) the \textit{out-of-bag score}. In regression models, the R2 coefficient measures how well the regression predictions estimate the real data points. An R2 coefficient of determination equal to 1.0 signals that the predictions of the regression model perfectly fit the data. The \textit{out-of-bag score} gives an estimate of the error rate of the training model for new data from the same distribution. The out-of-bag score is calculated by averaging only the trees for which a given data point prediction was not in the training data. This score is only available when \texttt{bootstrap=True}, which means that bootstrap samples are used when building the trees instead of using the whole data set for each tree. We enabled the out-of-bag score using the following command \texttt{oob\_score=True}.


\subsubsection{Features}
\label{subsubsec:features}

Our primary control feature is the rate constant for the chemical reactions. 
We consider four rate constants ($k$) of 1$\cdot10^{-9}$, 1$\cdot10^{-10}$, 1$\cdot10^{-11}$, 1$\cdot10^{-12}$ mol m$^{-2}$ s$^{-1}$. 
We partition the fracture-based features for our into three categories: (1) topological, (2) geometric, and (3) hydrological. 
All of the features we consider are fracture-based.

\paragraph{\textbf{Topological features}}
To obtain and measure the topological features of the DFN, we adopt a graph-based representation as presented in~\cite{hyman2018identifying}.
Therein the fractures are represented as nodes in the graph and intersections between fractures are represented as edges. 
If two fractures intersect in the DFN, then there is an edge in the graph between the corresponding nodes. 
Formally, let $\Omega = \{f_i\}$ for $i = 1,\ldots,N$ denote the entire network composed of $N$ fractures and let ${\ell_{i,j}}$ be the set of intersections between fractures, i.e., $f_i \cap f_j \neq \emptyset \equiv \ell_{i,j}$.
We define a graph $G = (V,E)$ of node set $V$ and edge set $E$ using a mapping $\phi$ defined in the following way. 
For every fracture $f_i \in \{f_i\}$, there is a node $v \in V$,
\begin{equation}
    \phi: f_i \rightarrow v_i\,.
\end{equation}
Similarly, for every intersection in ${\ell_{i,j}}$, there in edge $e_{i,j} \in E$ that connects the corresponding nodes in $V$
\begin{equation}
    \phi : f_i \cap f_j \neq \emptyset \rightarrow e_{i,j} = (v_i, v_j)\,.
\end{equation}
We also include nodes representing the inflow and outflow boundaries $G$ to account for the direction of flow.
Every fracture that intersects the inlet plane $\vx_0$ is connected to the source node $s$
\begin{equation}
    \phi:  f_i \cap \vx_0 \neq \emptyset \rightarrow (s, v_i)\,.
\end{equation}
Likewise, every fracture that intersects the outlet plane $\vx_L$ is connected to the source node $t$
\begin{equation}
    \phi:  f_i \cap \vx_L \neq \emptyset \rightarrow (v_i,t)\,.
\end{equation}
Similar mappings have been used in the literature~\cite{hope2015topological,andresen2013topology,hyman2017predictions}
Note that the mapping $\phi$ is an isomorphic bijection, which means that every subgraph in the graph corresponds to a unique subnetwork in the DFN. 
The graph is handled using \texttt{NetworkX}~\cite{hagberg-2008-exploring} and all graph-based features are computed using built-in algorithms within \texttt{NetworkX}.

\begin{figure}[t]
    \centering
    \includegraphics[width=0.9\textwidth]{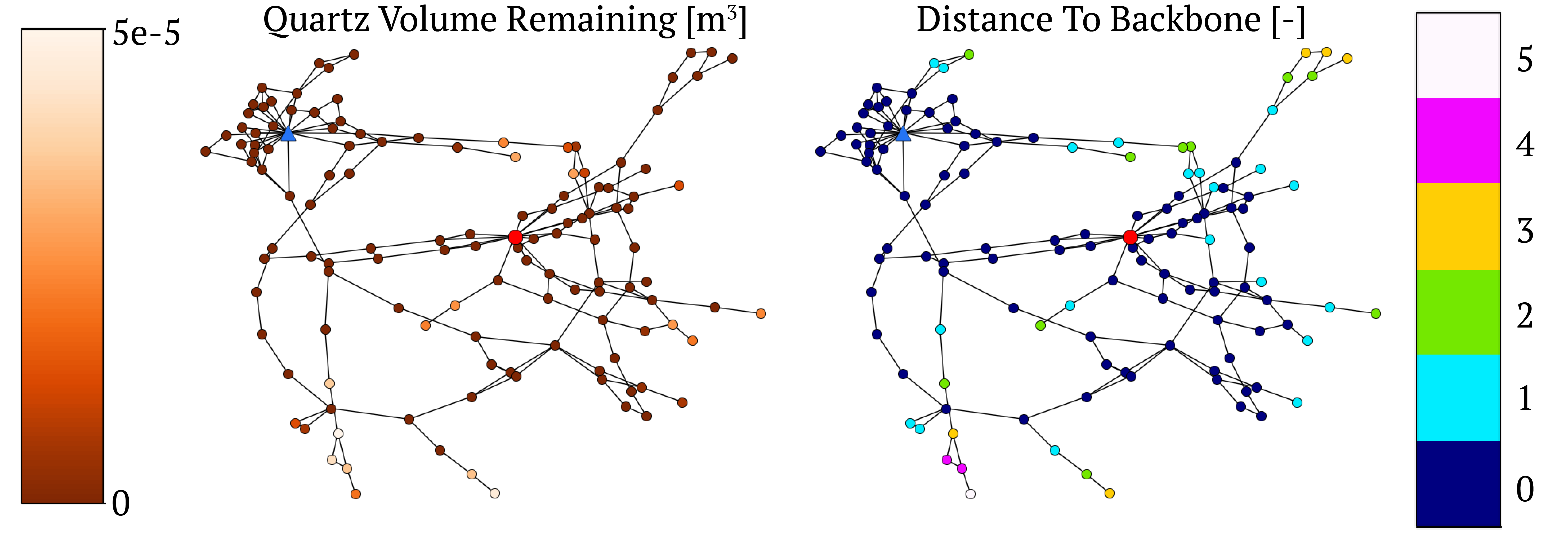} \\
    \caption{\label{fig:graphs} Graph representations based on the DFN shown above. (left) Nodes/fractures are colored by (left) quartz volume remaining (right) topological distance to the backbone. Distance to the backbone equal to 0 refers to a fracture on the backbone (primary sub-network membership).}
\end{figure}

Figure~\ref{fig:graphs} shows a graph representation of one DFN from our set of simulations, other networks exhibit the same behavior.
We represent each DFN using a graph to characterize topological structure influences for the distribution of reactive transport within the network.
In Figure~\ref{fig:graphs}~(a) fractures are colored by their remaining quartz volume at the end of the simulation and (b) fractures are partitioned by whether their effective P\'eclet number at the start of the simulation is diffusion dominated ($Pe < 1$) or advection dominated ($Pe \geq 1$).
There is almost no quartz remaining in fractures that were initially advection-dominated - compare Fig.~\ref{fig:graphs}~(a) and (b). 
However, there remains a substantial amount of quartz in the fractures that were initially diffusion-dominated. 
Flow channelization, isolated regions of higher volumetric flow rates, is a commonly observed phenomenon in field and laboratory experiments as well as numerical simulations in fractured media~\cite{abelin1991large,abelin1985final,dreuzy2012influence,frampton2011numerical,hyman2020flow,rasmuson1986radionuclide}. 
This phenomenon is observed in Figure~\ref{fig:graphs}~(b) where advection-dominated pathways connecting inflow and outflow boundaries (blue triangle and red circles respectively) can be seen. 
These flow channels indicate the existence of primary sub-networks, also referred to as  backbones, within the fracture network. 
To this end, we partition the network into disjoint primary and secondary sub-networks. 
Formally, let $\Omega = \{f_i\}$ for $i = 1,\ldots,N$ denote the entire network composed of $N$ fractures, then 
\begin{equation}
\Omega = \Omega^\prime \cup \Omega^\ast \qquad  \text{ and } \qquad  \Omega^\prime \cap \Omega^\ast  = \emptyset,
\end{equation}
where $\Omega^\prime$ and $\Omega^\ast$ are the primary and secondary sub-networks. 
We define the secondary network to be comprised of all dead-end structures, i.e., dead-end fractures and cycles, and the primary sub-network is its complement, similar to the methods proposed in \citet{doolaeghe2020graph} and \citet{yoon2023effects}.
Our partitioning definition is based solely on the network structure, specifically the topology/connectivity, and does not utilize any hydraulic, geometric, or flow information.
To do so, we compute the current-flow through the network using Kirchhoff's laws with one unit of current is injected into the graph at the source node and one unit is extracted at the target,  and every edge has unit resistance.  
Kirchhoff's law expresses that the sum of incoming currents at a node must be equal to the sum of outgoing currents.
Defining the sign of incoming currents as positive and the sign of outgoing current as negative, then the law can be represented
that the sum of the currents at each node is zero, 
\begin{equation}
    \sum_j^n I_{i,j} = 0~\forall~v_i~\in~V\,.
\end{equation}
Once the current is determined on every edge, all edges with current less than machine precision ($\epsilon = 10^{-16}$) are removed.
\begin{equation} 
E^\prime = \{e_{i,j}\} \text{ if } I_{i,j} > \epsilon\,.
\end{equation}
Then any nodes that are no longer connected to the sub-network containing the source are target are removed. 
\begin{equation}
    V^\prime = \{v_i\} \text{ if } |e_{i,\cdot}| > 0,  
\end{equation}
where $|e_{i,\cdot}|$ is the number of edges with an endpoint on the node $v_i$.
Then our primary sub-graph is defined as 
\begin{equation}
    G^\prime = (V^\prime, E^\prime),
\end{equation}
and the secondary its complement
\begin{equation}
    G^\ast = G \setminus G^\prime\,.
\end{equation}
Recalling that the mapping $\phi$ is a bijective isomorphism allows us to extract the primary and secondary sub-networks of the DFN 
\begin{equation}
    \phi^{-1}: G^\prime \rightarrow \Omega^\prime \qquad \text{and} \qquad \phi^{-1}: G^\ast \rightarrow \Omega^\ast.
\end{equation}

\begin{enumerate}
    \item \emph{Node Degree: } The degree of a node $v_i$ is the number of edges with one endpoint on the node $v_i$. It describes the number of intersections on the corresponding fracture. Formally, 
    \begin{equation}
        \text{Degree} (v_i) = \sum_{j=1}^n A_{ij},
    \end{equation}
    where $A_{i,j}$ is the adjacency matrix of the graph $G$ in which the entry $a_{i,j}$ is the number of edges in $G$ with end points $e_{i,j} = (v_i, v_j)$. Nodes with a larger degree are well connected and ought to be a location of substantial transport and reactions. 
    \item \emph{Degree Centrality: } Degree centrality is a normalized measure of the node degree defined above. For vertex $i$,
    \begin{equation}
    \label{eq:degree_cent}
    \text{Degree centrality}(v_i) = \frac{1}{n-1} \sum_{j=1}^n A_{ij}.
    \end{equation}
    High degree centrality indicates that a node is well connected and such nodes tend to be concentrated in the core of the network. Nodes with low degree centrality are often in the periphery or on branches that cannot conduct significant flow and transport. Physically, it describes the number of other fractures that intersect with a given fracture. 
    \item \emph{Distance to Backbone: } If a fracture is on the backbone, then the distance to the backbone is 0. If the fracture is in the secondary sub-network, then the value is the shortest topological path from that fracture to one on the backbone. Formally, 
    \begin{equation}
    \text{Distance to backbone} (v_i) = 
        \begin{cases}
            0, & \text{if } f_i =  \Omega^\prime\\
            \min d(f_i,f_j)~\forall~f_j~\in \Omega^\prime & \text{if } f_i \neq \Omega^\ast
        \end{cases}
    \end{equation}
    where the topological distance metric $d(u_i, u_j)$ counts the number of edges between the nodes $u_i$, $u_j$. In practice, we use an unweighted Dijkstra's method implemented in \texttt{NetworkX} for the search. 
    Primary/secondary membership is a good indicator of the final quartz volume in each fracture -- compare Fig.~\ref{fig:graphs}~(a) and (c).
    Almost all of the quartz has been dissolved out of the primary sub-network and there is a significant volume remaining in the secondary sub-network fractures. Beyond a Boolean of backbone membership, the distance to the backbone includes more information about the proximity of a fracture in the secondary sub-network to the backbone. Reactions in fractures that are farther away from the backbone tend to transport limited and therein remains more quartz. 
    \item \emph{Betweenness Centrality: } The betweenness centrality~\cite{anthonisse1971rush,freeman1977set} of a node describes the extent to which a node can control communication on a network. Consider a path with the fewest possible edges (geodesic path), that connects a node $u$ and a node $v$ on a graph. In general, there may be more than one such paths, and with $\sigma_{uv}$, we denote the number of such geodesic paths. Furthermore, let $\sigma_{uv}(i)$ denote the number of such paths that pass through node $i$. We then define, for node $i$,
    \begin{equation}
    \label{eq:betweeness}
    \text{Betweenness centrality} = \frac{1}{(n-1)(n-2)}\sum_{\substack{u,v = 1 \\ u\neq i\neq v }}^n \frac{\sigma_{uv}(i)}{\sigma_{uv}},
    \end{equation}
    where the leading factor normalizes the quantity so that it can be compared across graphs of different sizes $n$. Many backbone nodes have high betweenness values; however, other paths through the network can show high values, since this feature considers \emph{all} paths in the graphs, and not only those from source to target. 
    
    \item \emph{Source-to-target Current Flow: } This is a type of centrality measure adopted from an electrical current model~\cite{brandes2005centrality}. The current flow assumes a given source and target. Imagine that one unit of current is injected into the network at the source, one unit is extracted at the target, and every edge has one unit of resistance.
    Then, the current flow centrality is given by the current passing through a given node. This can be described by Kirchhoff's laws, or in terms of the graph Laplacian matrix $\mathbf{L}=\mathbf{D}-\mathbf{A}$, where $\mathbf{A}$ is the adjacency matrix for the graph and $\mathbf{D}$ is a diagonal matrix specifying node degree: $D_{ii}=\sum_j A_{ij}$.  We can define the current flow for node $i$ as
    \begin{equation}
    \label{eq:currentflowbetweeness}
    \text{Current flow }(v_i) = 
    \sum_{j=1}^n A_{ij} \bigl| \bigl(L_{is}^{+} - L_{js}^{+}\bigr) - \bigl(L_{it}^{+} - L_{jt}^{+}\bigr) \bigr|,
    \end{equation}
    where $\mathbf{L}^+$ is the Moore-Penrose pseudoinverse of $\mathbf{L}$, $s$ is the source node, and $t$ is the target node.

  The current flow centrality is often referred to as random-walk centrality~\cite{newman2005measure}, measuring how often a random walk from the source ($s$) to the target ($t$) passes through a node $i$. Unlike betweenness centrality, the current flow centrality considers only the paths from source to target, it's values are zero on any branch of the graph outside of the central core. Thus, we expect a correlation between high current flow values and nodes that have a large influence on the transport from source to target.
\end{enumerate}

\paragraph{\textbf{Geometric Features}}
The following features are based on the fracture geometry. Recall that fractures are planar discs with the same initial aperture and radius. So, initially, they all have the same surface area and volume.  However, if a fracture intersects with the domain boundary, its shape is modified to conform to the boundary. Therefore, there is variation in the following attributes.
\begin{enumerate}
    \item \emph{Surface area:} The surface area of the polygon representing the fracture plane. For a non-truncated fracture, the value is equal to $\pi r^2$. 
    \item \emph{Total fracture volume:} The total volume of the fracture is the surface area, which varies between fractures, and the aperture, which is initially constant. 
    \begin{equation}
    \label{eq:volume}
    \text{Total fracture volume }n_i = V_i\,.
    \end{equation}
    The total fracture volume determines the initial amount of quartz in each fracture, being the total volume multiplied by the initial volume fraction (80\%). Because the total quartz volume is a constant scaled quantity of the total fracture volume, we only include the total fracture volume as a feature. 
    \item {\emph{Projected volume: } The projected volume of the fracture is the component of a fracture's volume-oriented parallel to the main flow direction (inlet to outlet plane). 
    Assuming the flow is oriented along the $x$-axis, the projected volume is expressed as
    \begin{equation}
    \label{eq:volprojection}
    \text{Projected volume} = V_i \sqrt{(\mathbf{O}_i)_y^2 + (\mathbf{O}_i)_z^2},
    \end{equation}
    where $V_i$ is the volume of fracture $i$ and $\mathbf{O}_i$ is the unit vector normal to the fracture plane, called the orientation vector. Since the flow is oriented along the $x$-axis, the projected volume is expressed by the projection of $\mathbf{O}_i$ onto the $yz$-plane. Fractures that are oriented along the main flow direction are more likely to carry a significant part of the flow, compared to the fractures oriented perpendicular to the normal flow direction. 
    
    \item \emph{Intersection area: } The intersection area is the total length of intersections on a fracture multiplied by the initial aperture. 
}

\end{enumerate}

\paragraph{\textbf{Hydrological Features}}
The following are a set of hydrological features that are computed on a pipe-network representation of the DFN, cf.~\citet{karra2018modeling} for details about how the pipe/graph-network is obtained and the numerical simulations are performed. 
Steady pressure-driven flow is computed to obtain pressure and volumetric flow rates throughout the DFN. 
Obtaining these values is computationally inexpensive especially when compared to the high-fidelity reactive transport simulations. 
\begin{enumerate}
    \item \emph{Volumetric Flow Rate: } We compute the volumetric flow rate of fluid passing through each fracture. Given the flow rates into and out of the fracture, we take a single value that is one-half the absolute value of total flow exchanged by a fracture with its neighbors,
    \begin{equation}
         Q(f_i) = \frac{1}{2} \sum_j^n | Q_{i,j}|.
    \end{equation}
    The absolute value is necessary because of the sign dependence of flow into (positive) and outgoing (negative) and the 1/2 is to account for double counting. 
    \item \emph{\pe Number: }We compute the \pe number of each fracture using the volumetric flow rate $Q_i$, fracture radius $r_i$, and surface area $S_i$ and the diffusion coefficient $D = 10^{-12}\textrm{m}^2/\text{s}$
    \begin{equation}
        Pe(f_i) = \frac{Q_i r_i}{S_i D}.
    \end{equation}
    Regions with a high \pe tend to occur within the primary sub-network and reactions therein are kinetically limited, i.e., there is sufficient volumetric flow to flush away the aqueous silica, and once the quartz is fully dissolved in the primary sub-network, than a quasi-steady state is reached, and the overall apparent dissolution rate slows.
    \item \emph{Advective Damk\"ohler Number: }The advective Damk\"ohler number compares the reaction timescale to the convection. We compute it on a fracture basis, similar to the \pe number. First, we convert the rate constant $k$ [mol m$^2$/s] used in simulations to a rate $k^\prime$ [m/s] using the quartz molar volume (V$\text{m}$ = 22.6880 $10^{-6}$ m$^3$/mol, $k^\prime = k V\text{m}$). Then our advective \da number is defined as 
    \begin{equation}
        Da_I(f_i) = \frac{k S_i}{Q_i}.
    \end{equation}
    \item \emph{Diffusive Damk\"ohler Number: }Likewise, we compute the diffusive \da which compares the reaction timescale to diffusion and is given as
    \begin{equation}
        Da_{II}(f_i) = \frac{k r_i}{D}.
    \end{equation}
    
\end{enumerate}
In the following section, we discuss the feature importance analysis obtained from the random forest regression model, as well as the accuracy of the trained ML models. 

\section{Results}
\label{sec:results}
We used 800 generic networks for training and testing the ML random forest regression model. Each DFN was run with four reaction rate constants, ($k$) of 1$\cdot10^{-9}$, 1$\cdot10^{-10}$, 1$\cdot10^{-11}$, 1$\cdot10^{-12}$ mol m$^{-2}$ s$^{-1}$, equal to a total of 3256 networks. The DFNs consist of a set of fractures (on average 133 fractures per network) and each fracture is considered a sample and was used to train or test the random regression model. The total number of fractures in the data set is 446,129 fractures, and two-thirds (2/3) of the data were used for training the ML model (\textit{training set}), while the rest of the data (1/3) were used for testing the model (\textit{testing set}). We trained a regression model to predict the quartz volume that remains in a given fracture depending on three different attribute sets -- using only topological, geometric, and hydrological features as described above, along with the rate constant which we consider a primary control feature. Our results show that having access to flow features enhances the accuracy of the quartz volume prediction, however, it is not significant. 

In the following subsection, we discuss the feature importance analysis, giving us insights into the significance of each feature in predicting the remaining quartz volume within a given fracture. This analysis shows an interplay between the feature categories and their correlations. Later, we show details on the performance of the following two types of random forest regression models: (1) including all rate constants and gradually adding more complexity regarding the feature categories. For this type, we generated the following three regression models: \textbf{RF-1} model using only topological features; \textbf{RF-2} model using topological and geometric features; \textbf{RF-3} model using topological, geometric, and hydrological features; (2) including one rate constant at a time while using all features, resulting in four regression models for each of the rate constants.

\subsection{Feature Importance}
\begin{figure}[h]
    \captionsetup[subfigure]{labelformat=empty,position=top, labelfont=bf,textfont=normalfont,singlelinecheck=off,justification=raggedright,labelsep=space,indention=0pt,margin=0pt,skip=-2pt,hypcap=false}

    \centering
    \subfloat[\textbf{RF-1}]{{\includegraphics[scale=0.55]{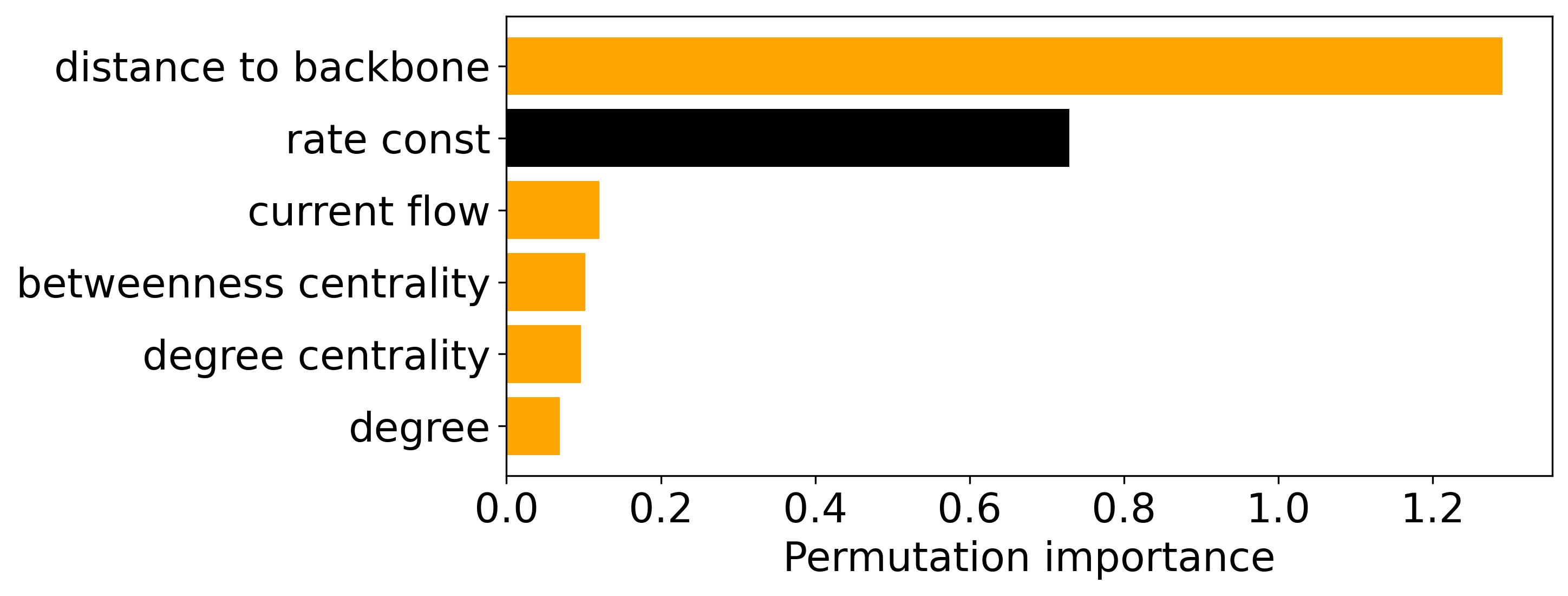} }}\\
    \vspace{-0.6cm}

    \subfloat[\textbf{RF-2}]{{\includegraphics[scale=0.55]{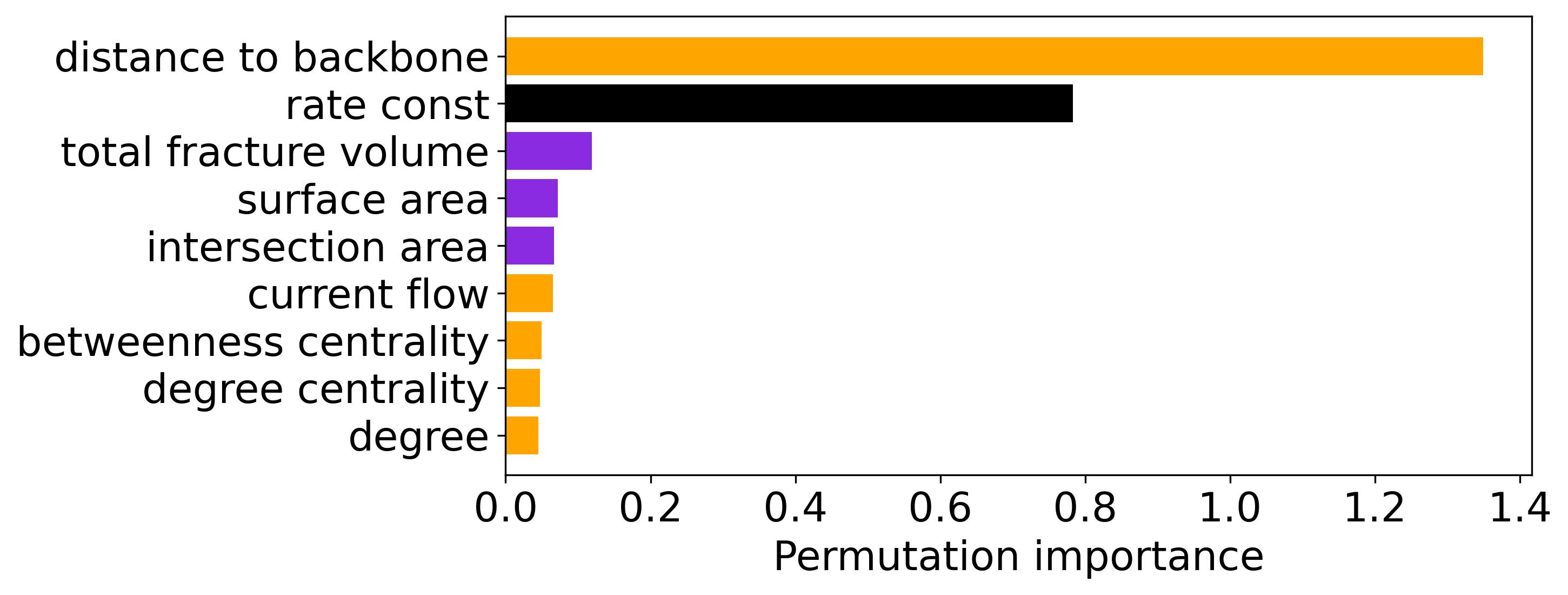} }}\\
    \vspace{-0.6cm}
    
    \subfloat[\textbf{RF-3}]{{\includegraphics[scale=0.55]{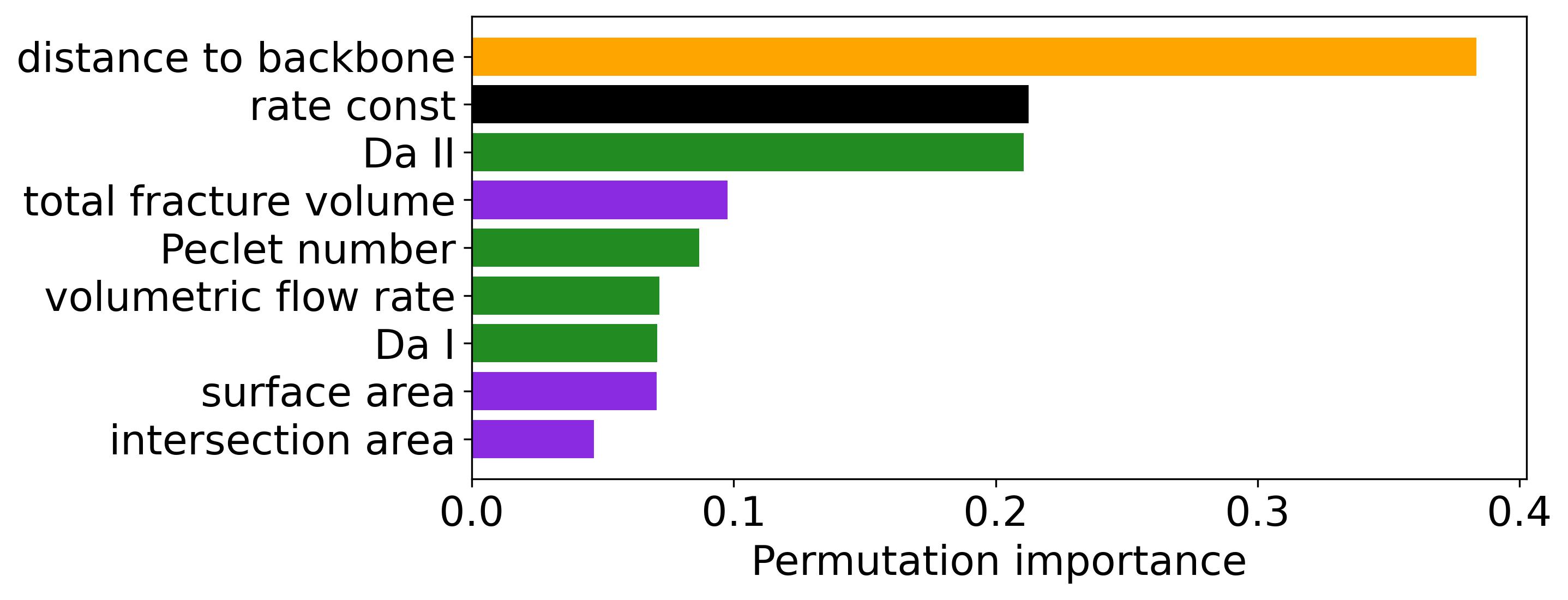} }}
    
    \caption{Feature importance analysis for three regression models: RF-1 (topological features), RF-2 (topological and geometric features), and RF-3 (hydrological, topological and geometric features). The topological features are depicted in orange, the geometric features in purple, the topological features in green, and the reaction rate constant in black. The importance analysis for all models show that the distance to the backbone and the rate constant are the most important features in our study.}
    \label{fig:feature_importance_all_rc}%
\end{figure}

As mentioned previously the random forest regression model can be used to identify the importance of each individual feature for the performance of the trees. Figure \ref{fig:feature_importance_all_rc} depicts the input feature importance analysis results for the three random forest regression models (RF-1, RF-2, and RF-3), with a gradually increasing number of features. Each model uses all values of the rate constant, shown in black. The topological features are displayed in orange, the geometrical features in purple, and the hydrological features in green. 

The feature importance analysis is calculated by comparing the baseline model to the model obtained by permutating the feature column. As shown in Figure~\ref{fig:feature_importance_all_rc}, the analysis confirms that the distance to the backbone and the rate constant are the two main quantities controlling the dissolution of quartz in the fractures for all regression models. This is not surprising since the distance to the backbone indicates how far a fracture is with respect to the backbone. The transport in fractures that are farther away from the backbone tend to be diffusion-dominated, thus therein remains more quartz. Contrary, when a given fracture is part or in the vicinity of the primary sub-network, the behavior in the fracture is advection-dominated, which leads to the complete depletion of quartz within the fractures. The rate constant is the second important feature for all trained regression models because it indicates the strength of dissolution in the fracture. There is a strong correlation between the distance to the backbone, reaction rate, and the quartz volume that remained in the system, since as the reaction rate increases also the dissolution penetrating deeper levels of the secondary fracture sub-network. This means that as higher the rate constant is, the more quartz will be flushed out from the secondary sub-network (taking into account that all quartz is dissolved from the primary sub-network). Conversely, when the rate constant is small, the reactive transport is reduced almost entirely to the primary sub-network.

As we increase the number of features from topological to topological and geometric, we can see that the total fracture volume, surface are and intersection area line up right after the first two topological features. When we add hydrological features, we observe that the diffusive Damk\"ohler number (Da$_{II}$) becomes almost as important as the rate constant, followed by the total fracture volume and other hydrological features. From the training results shown in Table~\ref{tab:grid_search_results}, we know that the model accuracy does not improve drastically, when more features are included in the training; however, from the features importance analysis, we can confirm that they carry significant information for the flow and reactive transport of a fracture.
Reactions during early simulation times are kinetically limited occurring mostly in the primary network. 
However, once all of the quartz in the primary network is dissolved, the secondary network becomes the location where all reactions are occurring. 
Flow in the secondary network is slow compared to the primary network (P\'eclet number $< 1$) and reactions therein are transport limited rather than kinetically. 
Thus, reactions in the secondary network are a balance between diffusion and the reaction rate, which is diffusive  Damk\"ohler number measures.
In other words, Da$_{II}$ captures the integrated effects of the distance to the backbone, rate constant, and P\'eclet number into a single variable which is then linked with late time dissolution effects and prediction of the remaining quartz in the system. 

\begin{figure}[!t]
    \captionsetup[subfigure]{labelformat=empty,position=top, labelfont=bf,textfont=normalfont,singlelinecheck=off,justification=raggedright,labelsep=space,indention=0pt,margin=0pt,skip=-2pt,hypcap=false}
    \centering
    \subfloat[\textbf{RF-2}]{{\includegraphics[scale=0.22]{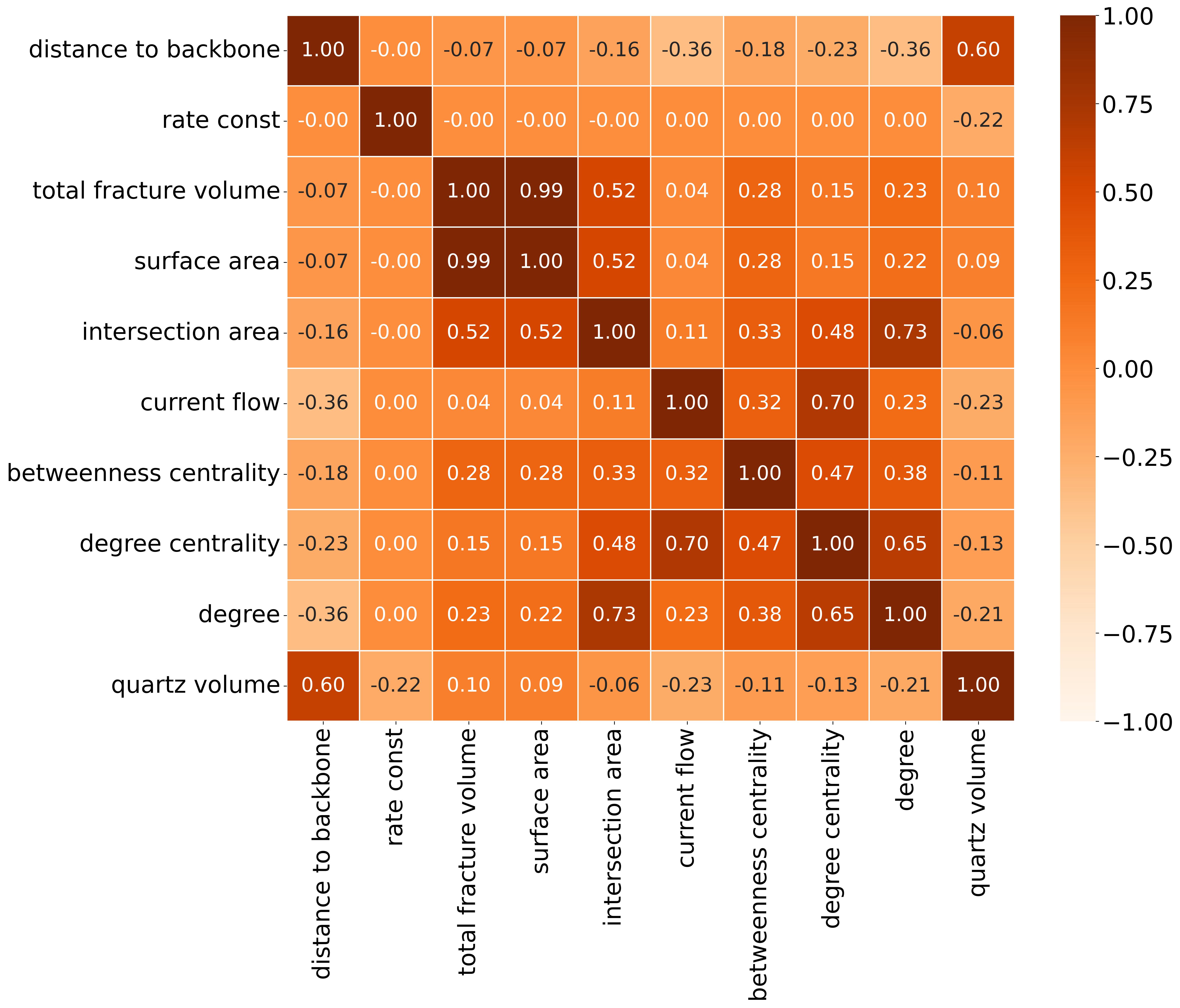} }}\\
    \vspace{-0.6cm}
    \subfloat[\textbf{RF-3}]{{\includegraphics[scale=0.22]{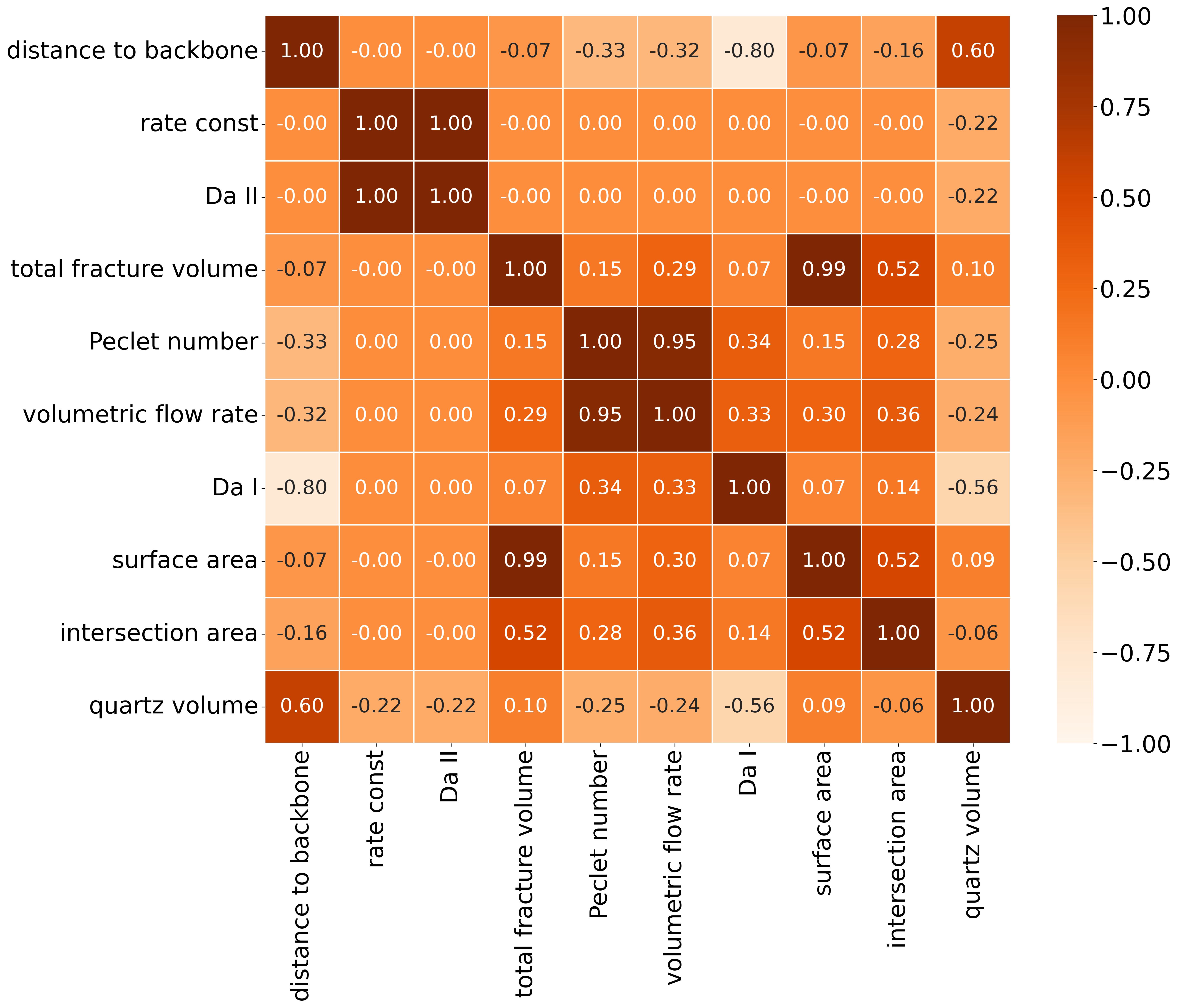} }}
    \vspace{-0.1cm}
    \caption{Correlation matrix of the model features displayed as a heat map for the remaining quartz volume and the first nine features used for the random forest regression training. The results for RF-2 show strong correlations between the following features: distance to the backbone and the quartz volume, total fracture volume and surface area, intersection area and degree, current flow, and degree centrality. When we add the hydrological features, some of the less significant topological features are not displayed. The RF-3 model shows a strong correlation between the aforementioned features and the following hydrological features: Da$_{II}$ and rate constant, volumetric flow rate and P\'eclet number. We observe a strong inverse correlation between Da$_{I}$ and the distance to the backbone, and Da$_{I}$ and quartz volume. The RF-1 model is omitted for clarity.}
    \label{fig:correlation_matrix}%
\end{figure}


Figure~\ref{fig:correlation_matrix} shows the correlations between input features (correlation matrix) in our regression model as a heat map. We omitted the correlation matrix for the RF-1 model since the topological features are depicted in RF-2 and RF-3. We reduced the number of features for each heat map to ten to improve the readability of the results. 

The results for RF-2 show strong correlations between the following features: distance to the backbone and the quartz volume, total fracture volume and surface area (0.99), intersection area and degree (0.73), current flow and degree centrality (0.70). When we add the hydrological features, some of the less significant topological features are not displayed. The RF-3 model show strong correlation between the aforementioned features and the following hydrological features: diffusive Damk\"ohler number (Da$_{II}$) and rate constant (1.0), volumetric flow rate and P\'eclet number (0.95). We observe a strong inverse correlation between the advective Damk\"ohler number (Da$_{I}$) and the distance to the backbone (-0.8), and the advective Damk\"ohler number (Da$_{I}$) and the quartz volume (-0.56). This is expected since these variables are derived from one another, and it confirms that the regression models learn the correct behavior by being able to predict those strong correlations. 

\subsection{Regression Models}
To implement the random forest regression model we used the \texttt{RandomForestRegressor} function from the \texttt{scikit-learn} module in Python. We performed an exhaustive cross-validated grid search over a number of parameter values using the \texttt{GridSearchCV} function. The optimized model was obtained after a grid-search through the following parameters: \texttt{n\_estimators}, \texttt{max\_depth}, \texttt{max\_features}, \texttt{min\_samples\_leaf}, and \texttt{min\_samples\_split}. The ranges for each of these hyperparameters are given in Table~\ref{tab:grid_search_values}. The input parameters used for the grid-search study include all the features: topological, geometric, hydrological, and the reaction rate constant (RF-3) as described in Section~\ref{subsubsec:features}. 

\begin{table}[H]
    \centering
    \begin{tabular}{|c|c|c|c|c|}
        \hline
        \textbf{RF Input Parameters} & \textbf{Grid Search} & \textbf{Base Model}    & \textbf{Optimized Model}\\
         \hline
          \multirow{2}{*}{\texttt{n\_estimators}}   & [10, 50, 100, 200, 300, 400,                          & \multirow{2}{*}{10}   & \multirow{2}{*}{1000} \\
                                                    & 500, 600, 700, 800, 900, 1000]                        &                       &                       \\ [.3em]
          \texttt{max\_depth}                       & [None, 30, 60, 90]                                    & None                  & 30                    \\ [.3em]
          \texttt{max\_features}                    & [None, sqrt, log]                                     & None                  & sqrt                  \\ [.3em]
          \texttt{min\_samples\_leaf}               & [2, 3, 4, 5]                                          & 2                     & 2                     \\ [.3em]
          \texttt{min\_samples\_split}              & [1, 2, 3, 4, 5, 6, 7, 8]                              & 1                     & 2                     \\ [.3em]
          \hline     
    \end{tabular}
    \caption{Hyperparameter ranges for the cross-validate grid-search and results. We show the default parameters used for the base model, as well as the optimized hyperparameters, resulting from the grid-search. To obtain these results, we use the RF-3 model, which contains all input features (topological, geometric, hydrological, and the reaction rate constant). Additionally, we set the following parameters for all regression models: \texttt{bootstrap=True},
    and \texttt{oob\_score=True}.} 
    \label{tab:grid_search_values}
\end{table}

The time needed for the cross-validated grid search was about 20 hours on a CPU machine. The time to train the R3 base model is approximately 20 seconds, while the optimized model needs around 7.5 minutes. This is mainly due to the number of estimators used for the optimized model. Even though the optimized model takes almost 22 times slower, the model is more robust and prone to overfitting. 

Table~\ref{tab:grid_search_results} shows a performance comparison between the base model (using the default parameters for the \\\texttt{RandomForestRegressor} function) and the optimized model using the R2 value obtained during training and testing, and the out-of-bag score. 

\begin{table}[H]
    \centering
    \begin{tabular}{|c||c|c||c|}
        \hline
          \multirow{2}{*}{\textbf{Model}}  &\multicolumn{2}{c||}{\textbf{R2 Values}} & \multirow{2}{*}{\textbf{OOB Score}} \\ \cline{2-3}
          
                                                              & \textbf{Train} & \textbf{Test}  &  \textbf{} \\
          \hline
          \textbf{Base RF-1}                                  & 0.9410  & 0.6883  & 0.6256     \\[0.2em]
          \textbf{Optimized RF-1}                             & 0.8864  & 0.7195  & 0.7179     \\[0.2em]\hline
          \textbf{Base RF-2}                                  & 0.9672  & 0.8193  & 0.7658     \\[0.2em] 
          \textbf{Optimized RF-2}                             & 0.9506  & 0.8397  & 0.8356     \\[0.2em]\hline
          \textbf{Base RF-3}                                  & 0.9701  & 0.8343  & 0.7845     \\[0.2em]
          \textbf{Optimized RF-3}                             & 0.9499  & 0.8509  & 0.8474     \\[0.2em]
          \hline
    \end{tabular}
    \caption{Performance measures of the random forest regression models. Comparison between the base regression model, using the default parameter values, and the optimized model, obtained from an exhaustive cross-validated grid search. These random forest models use all rate constants, while the number of input features is increased gradually with each model. The optimized model performs better than the base model. The out-of-bag score confirms that the optimized model is more robust. }
    \label{tab:grid_search_results}
\end{table}


The performance measured in Table~\ref{tab:grid_search_results} shows that the remaining quartz volume is hard to predict. When using only the topological features (RF-1 model), the random forest regression models (base and optimized) perform poorly, indicating that the topological features do not carry enough information to train the model. Adding geometric features enhances significantly the performance of both models, the base and the optimized. The R2 train values of the optimized model actually decrease in comparison to the base model; however, we expect that is due to overfitting of the base model. Including hydrological features (RF-3 model) only slightly increases the accuracy of the models, pointing to the complexity of the DFN system exhibiting dissolution. 

The overall accuracy of the training for the base and the optimized model is good, however, during testing the coefficient of determination (R2 value) drops significantly. This is very clear for the RF-1 base model, where only topological input features are used. The train R2 is $0.9544$, while during testing the model's test R2 value is only $0.7080$, which implies that the random forest regression models overfit during training and show higher accuracy in training while under-performing during testing. The optimized models overfit less and perform slightly better than the base model. The small difference between the test R2 values of the base and the optimized model suggests that performing a grid search and tuning the model parameters, even though important, it might not improve the accuracy of the training significantly. This strongly depends on the data set and the problem at hand. 

The out-of-bag score of the optimized model is eight to nine points larger than the one of the base model. This confirms that the optimized model is more prone to overfitting and is more robust.

\subsubsection{Regression Models Using All Rate Constants}
\begin{figure}[h]
    \captionsetup[subfigure]{labelformat=empty}
    \centering
    \small{\textbf{Train Results}}\vspace{-0.3cm}\par
    \subfloat{{\includegraphics[scale=0.4]{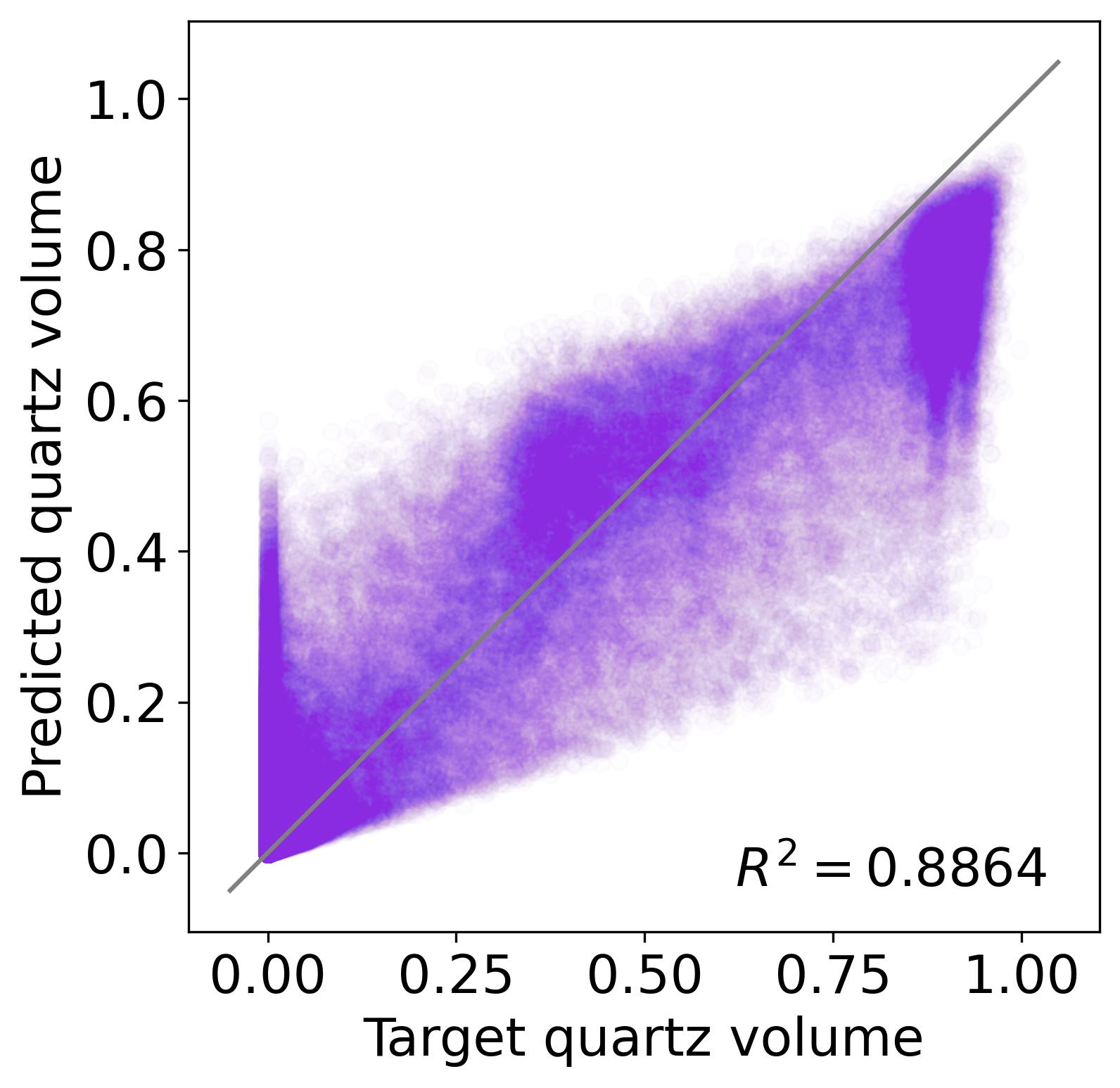} }}%
    \subfloat{{\includegraphics[scale=0.4]{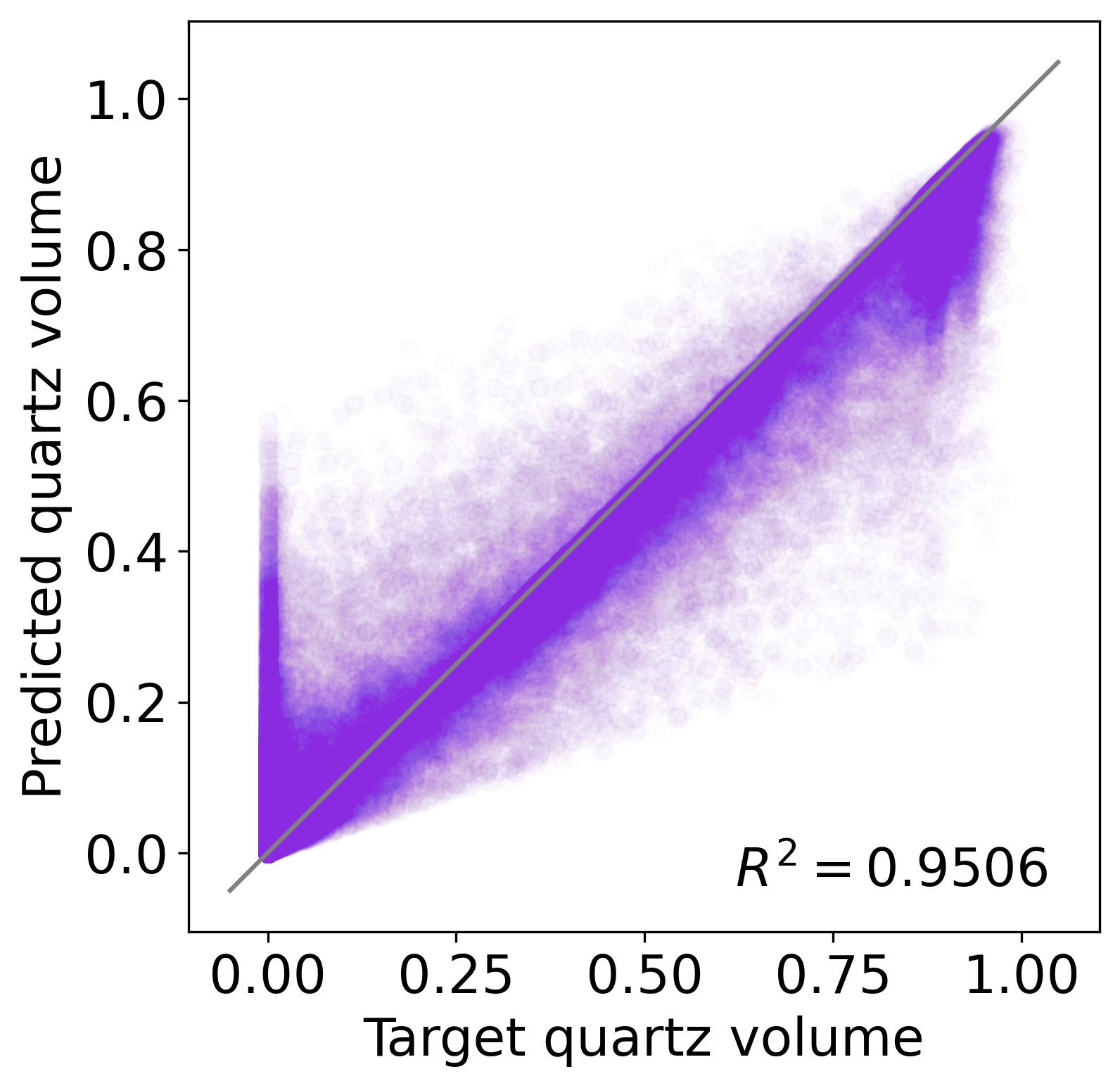} }}%
    \subfloat{{\includegraphics[scale=0.4]{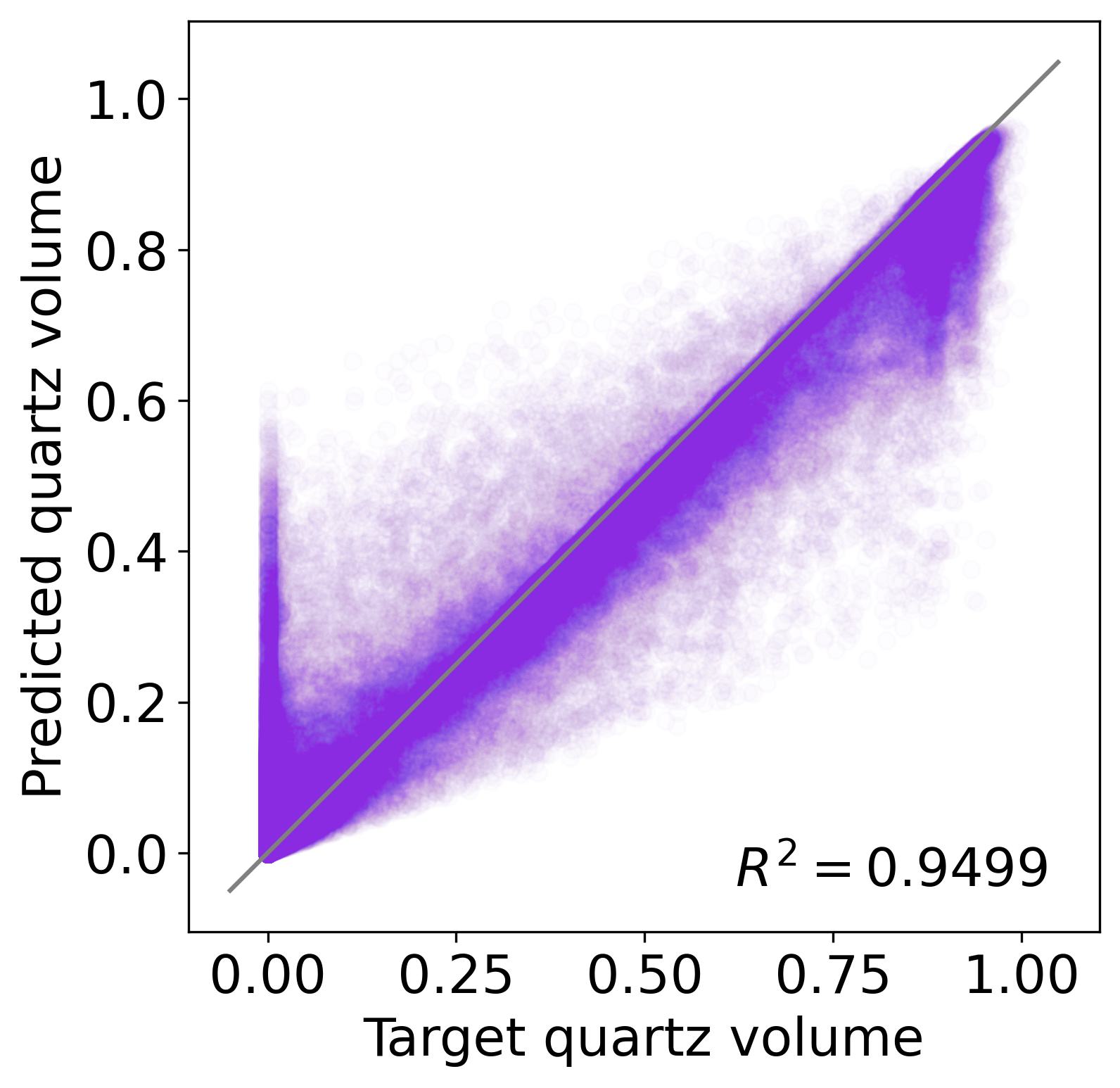} }}\\ 
    \vspace{0.2cm}
    \small{\textbf{Test Results}}\vspace{-0.3cm}\par
    \subfloat[\textbf{RF-1}]{{\includegraphics[scale=0.4]{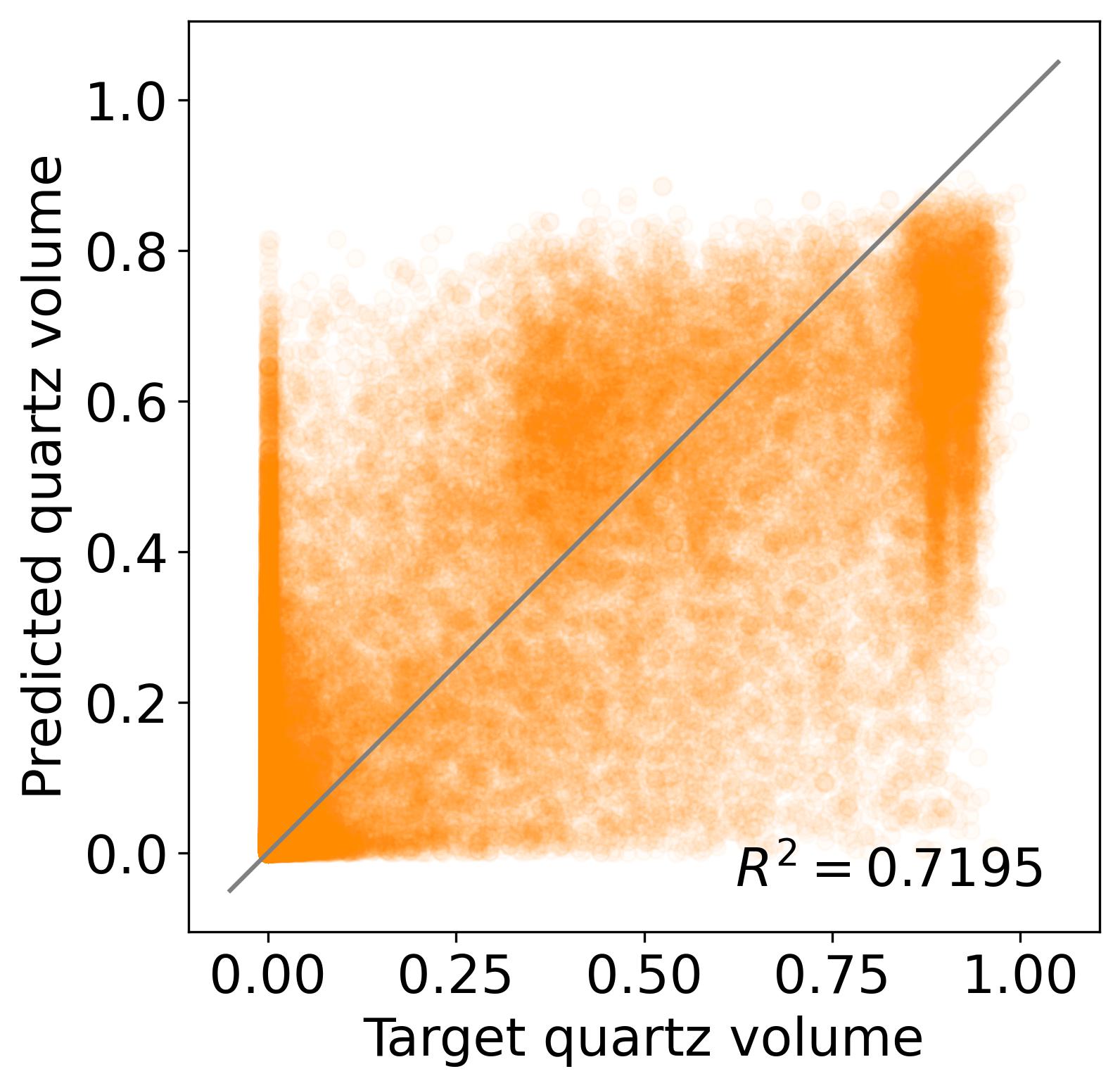} }}
    \subfloat[\textbf{RF-2}]{{\includegraphics[scale=0.4]{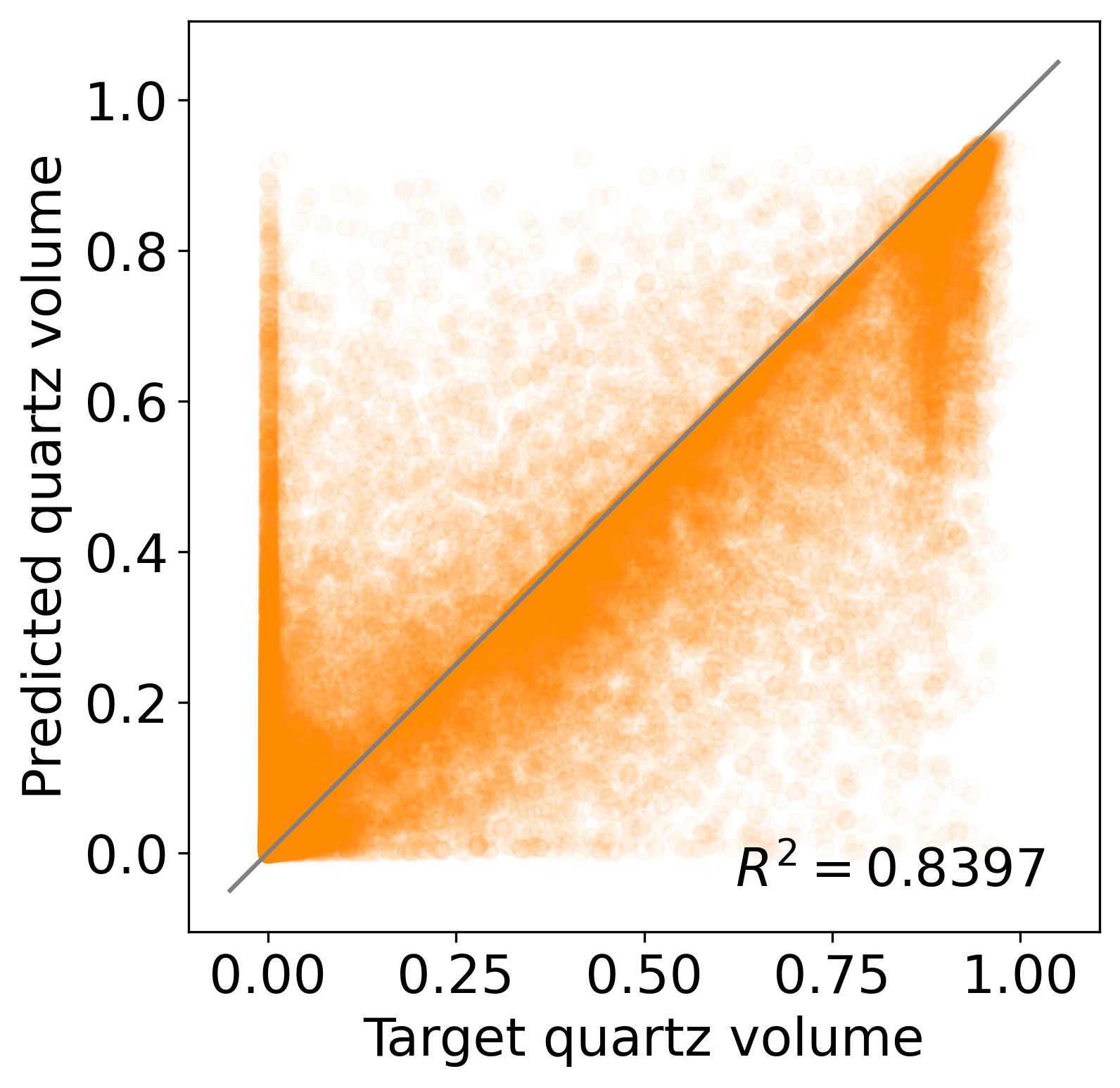} }}
    \subfloat[\textbf{RF-3}]{{\includegraphics[scale=0.4]{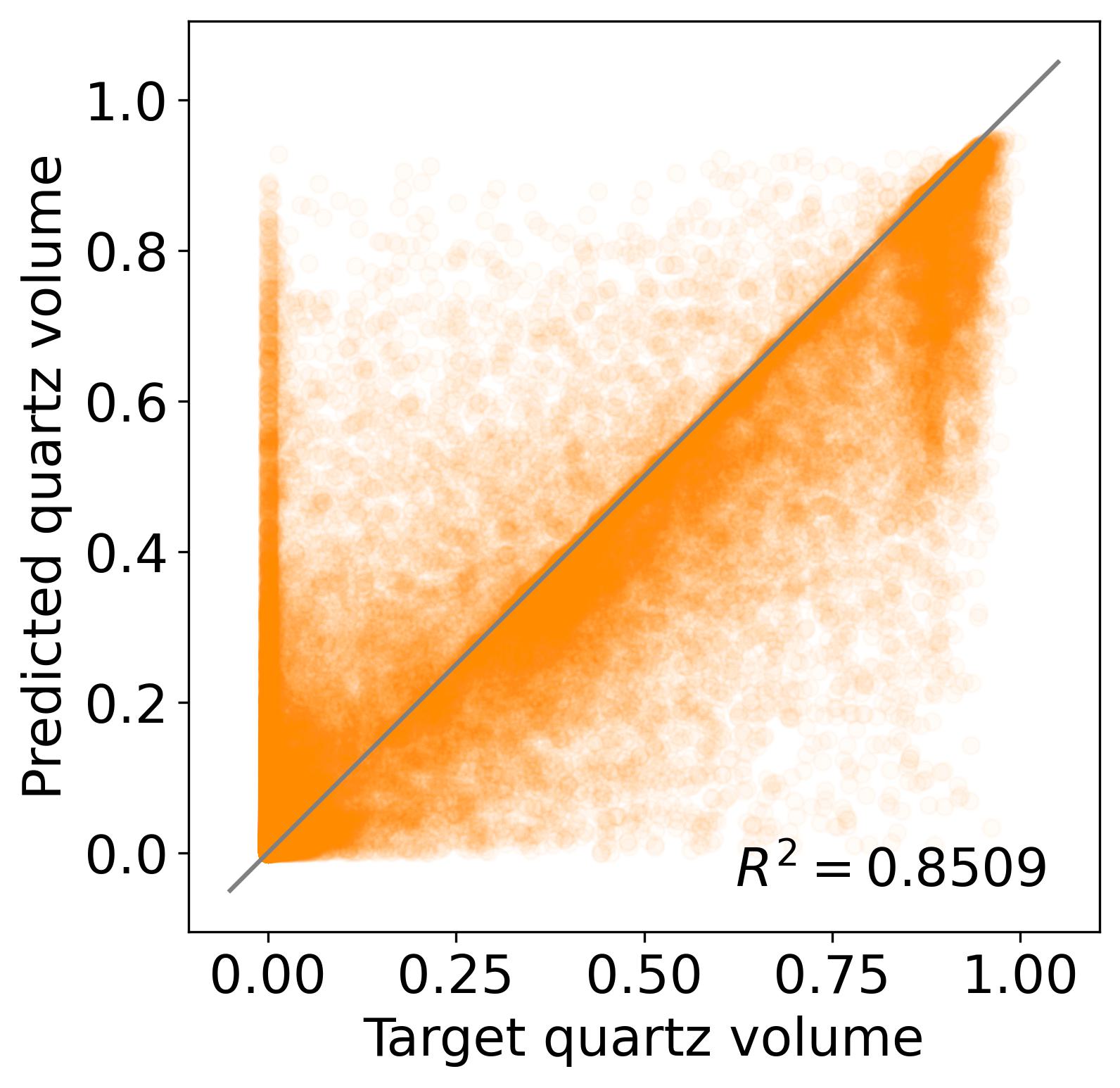} }}
    \caption{Random forest train (purple) and test (orange) predictions for the following models: RF-1 (topological features), RF-2 (topological and geometric features), and RF-3 (hydrological, topological, and geometric features). The RF-1 model does not include enough input parameters to predict correctly the remaining quartz volume in a given fracture. RF-2 shows tighter predictions; however, it still fails to predict the quartz volume for fractures that are either fully depleted or clogged (values close to 0 or 1, respectively). RF-3 gives slightly better predictions (see the R2 value); however, the improvement is not significant.}%
    \label{fig:rate_const_all}%
\end{figure}

Training the regression model on all rate constants allows for comparison between different feature categories. Figure~\ref{fig:rate_const_all} shows the train and test results for the RF-1, RF-2, and RF-3 models. 
We see that using only the topological features is not sufficient to obtain a good regression model. The RF-1 test results are scattered and the model is not able to predict the correct target quartz values. By adding the geometric features (RF-2 model), we observe much better predictions of the remaining quartz volume. However, the random forest regression model still has difficulties predicting the quartz volume, especially when the quartz is either fully depleted or clogs the fracture (normalized quartz volume close to 0 or 1) in advection- or diffusion-dominated regions, respectively. We trained a third model (RF-3) that includes the hydrological features, which performs slightly better than the RF-2 model; however, the improvement is not significant. 

\subsubsection{Regression Models Using A Single Rate Constant}
\begin{figure}[!ht]
    \captionsetup[subfigure]{labelformat=empty}
    \centering
    \small{\textbf{Train Results}}\vspace{-0.3cm}\par
    \subfloat{{\includegraphics[scale=0.3]{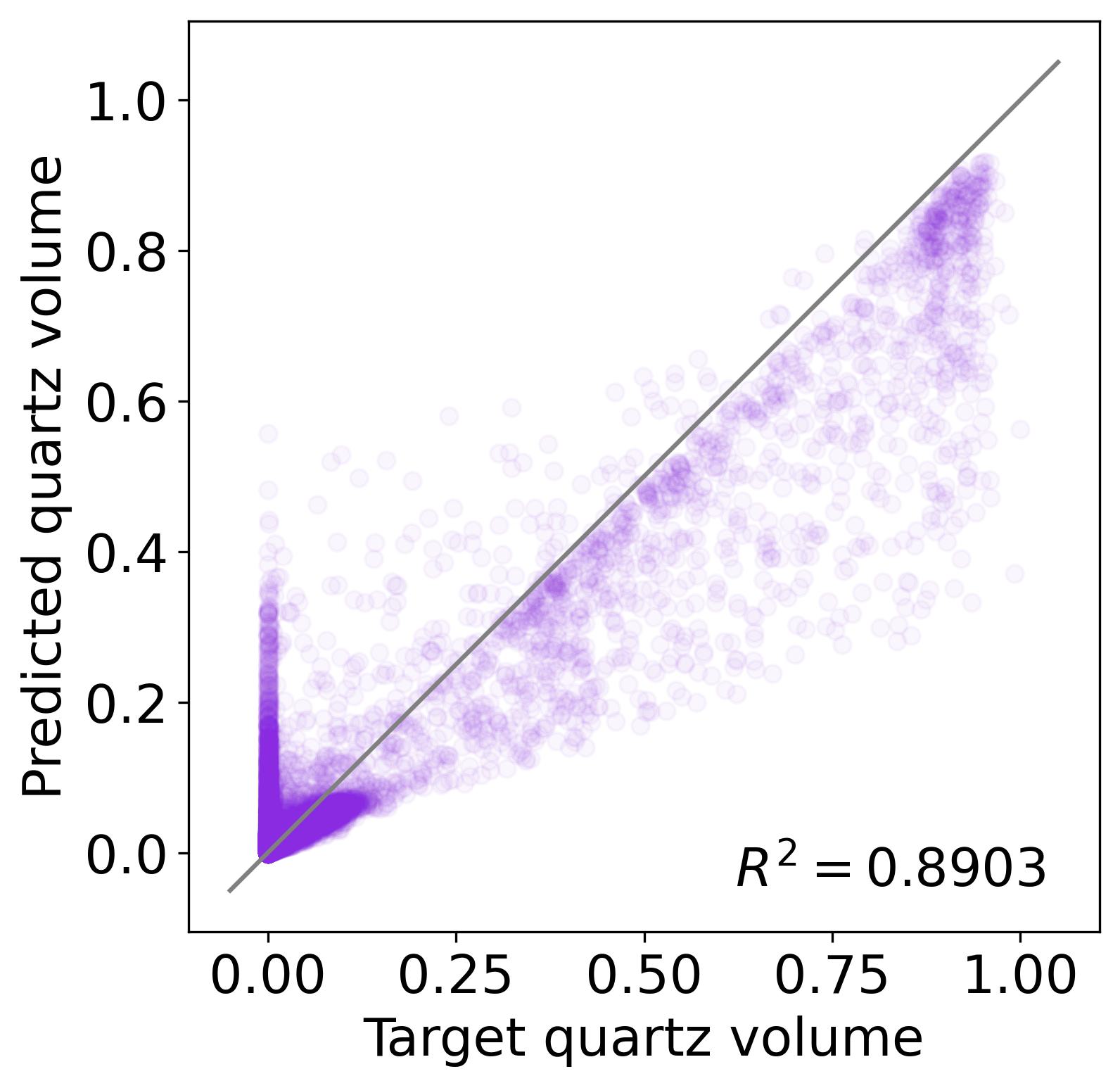} }}
    \subfloat{{\includegraphics[scale=0.3]{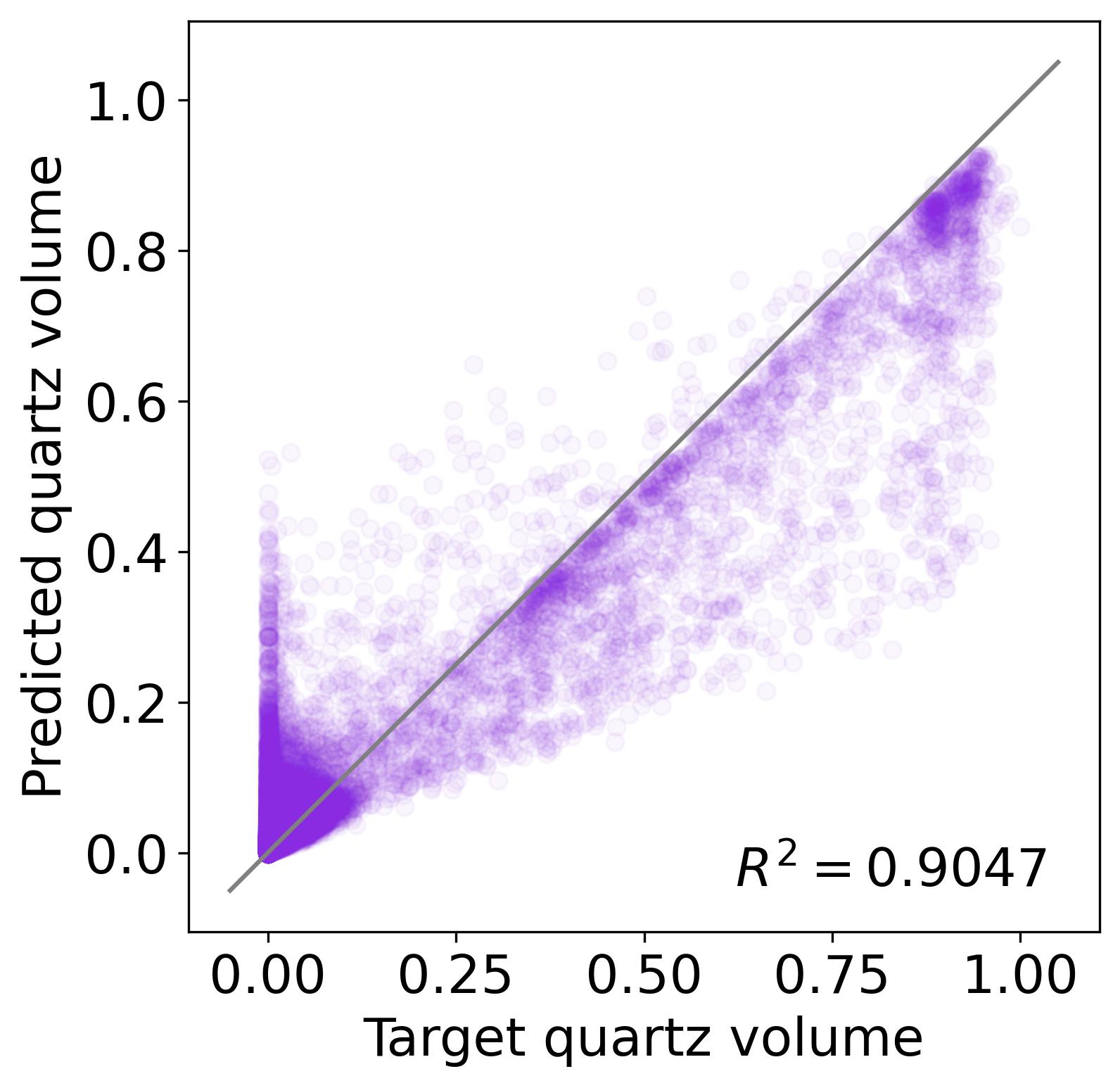} }}
    \subfloat{{\includegraphics[scale=0.3]{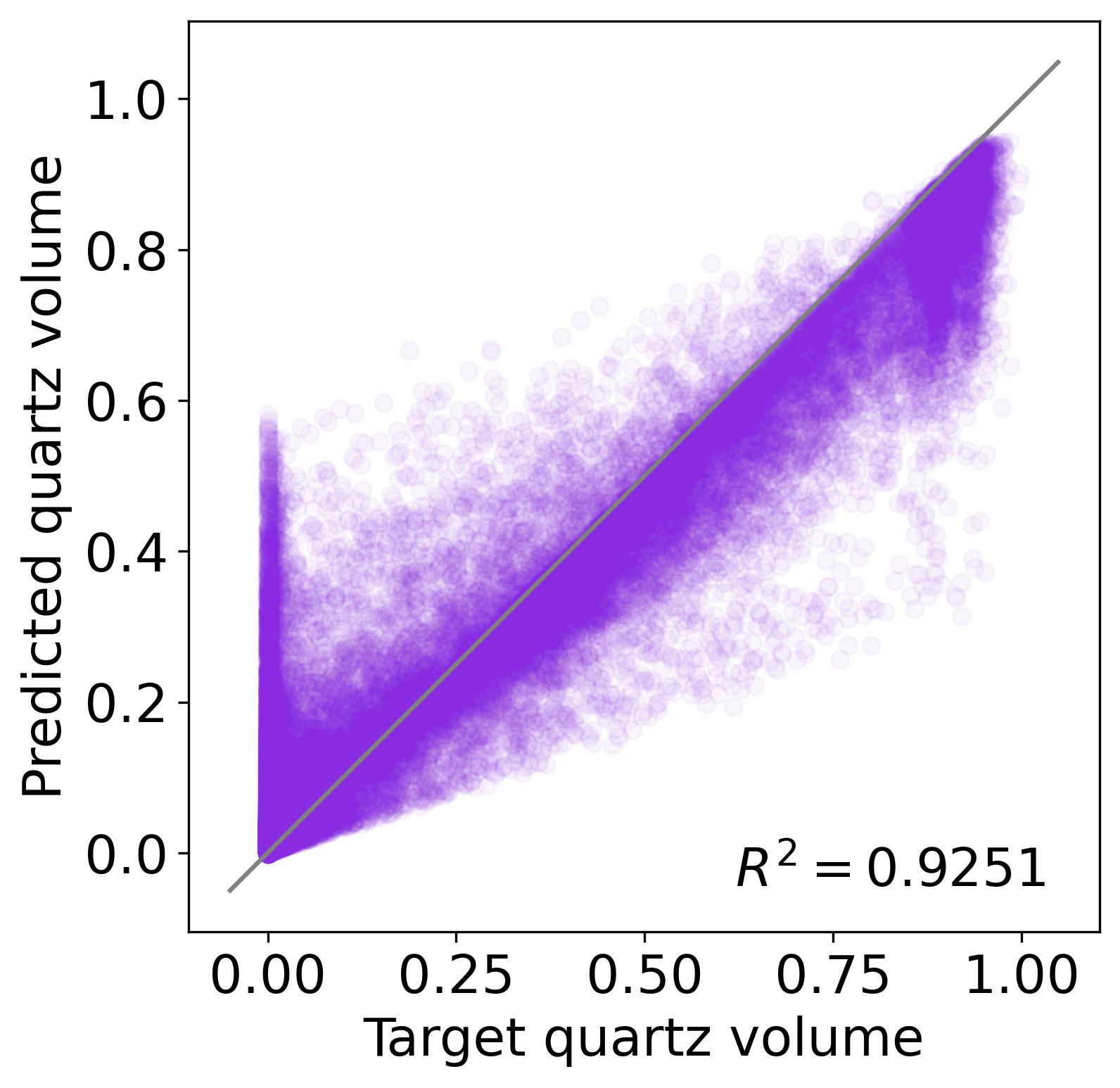} }}
    \subfloat{{\includegraphics[scale=0.3]{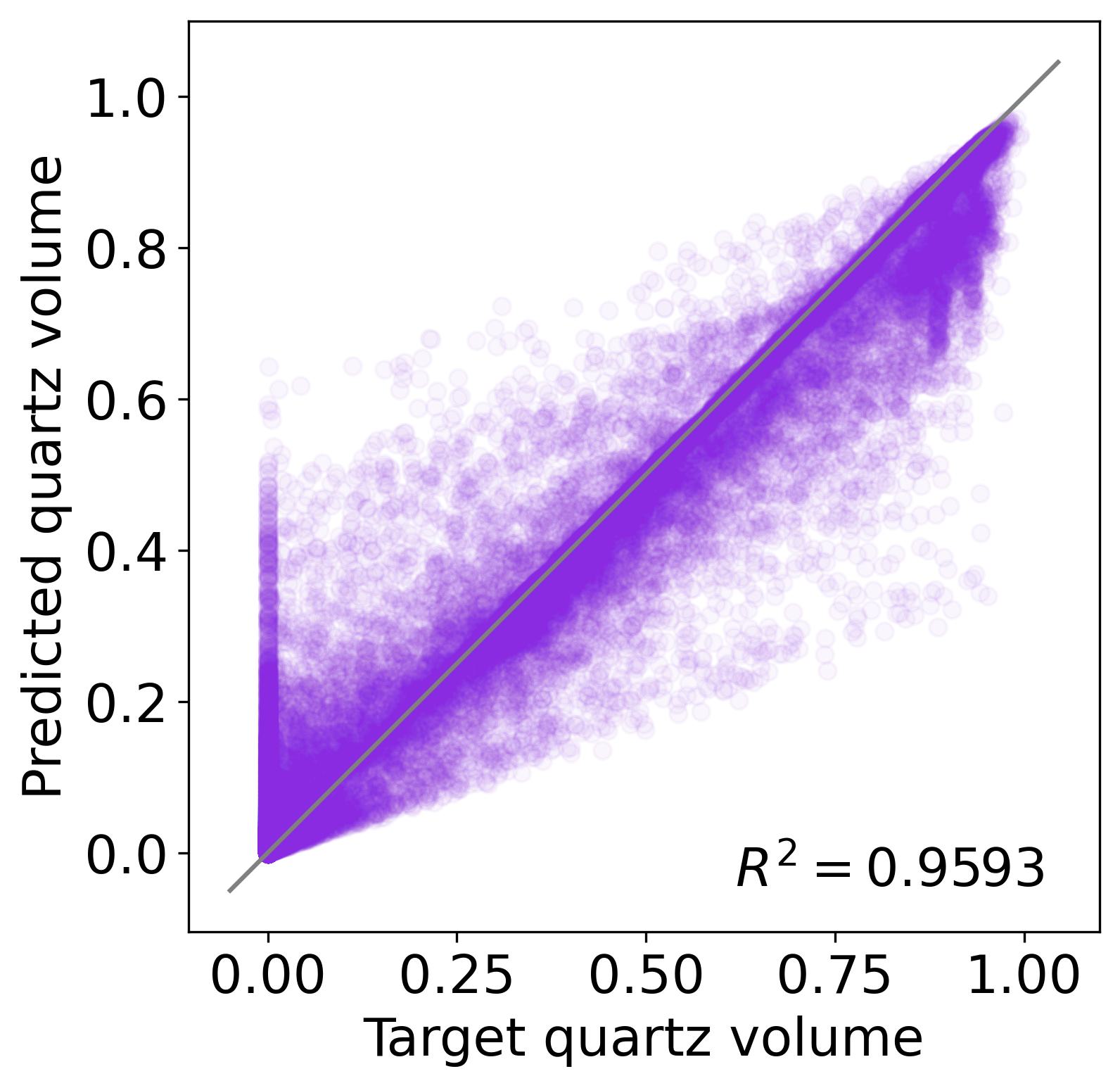} }}\\
    
    \small{\textbf{Test Results}}\vspace{-0.2cm}\par
    \subfloat[\textbf{(a)}\quad k = 1$\cdot10^{-9}$]{{\includegraphics[scale=0.3]{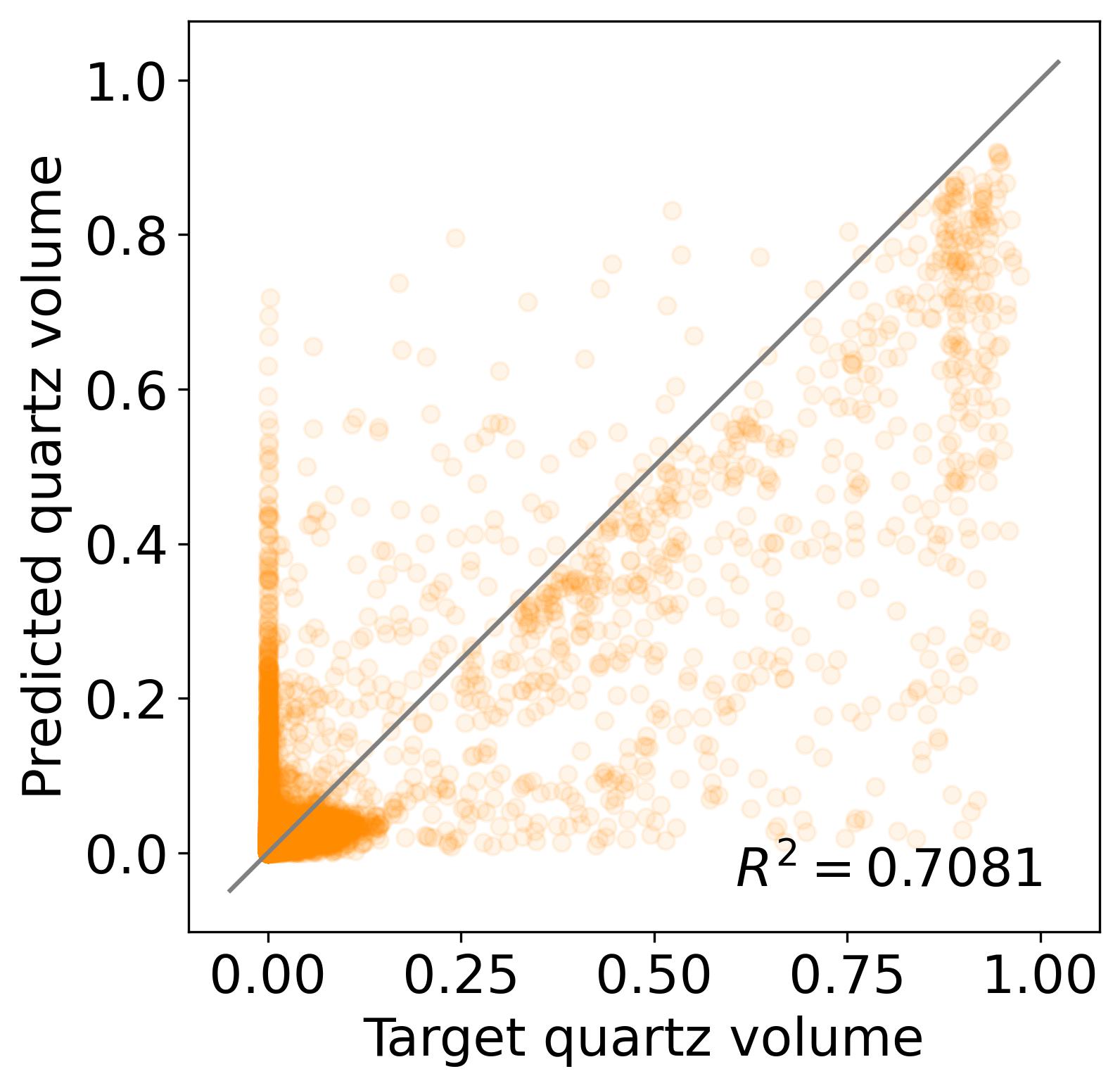} }}
    \subfloat[\textbf{(b)}\quad k = 1$\cdot10^{-10}$]{{\includegraphics[scale=0.3]{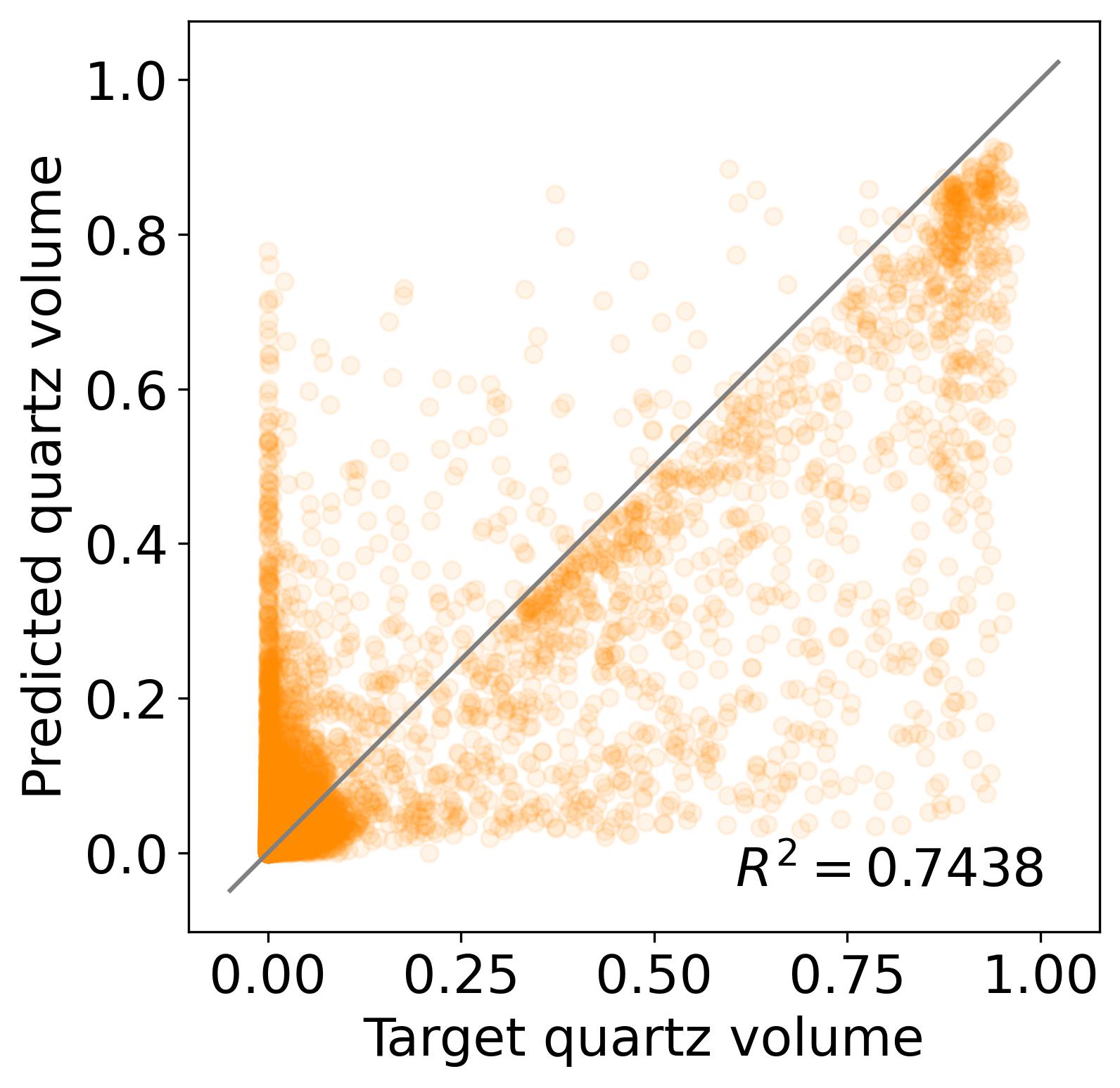} }}
    \subfloat[\textbf{(c)}\quad k = 1$\cdot10^{-11}$]{{\includegraphics[scale=0.3]{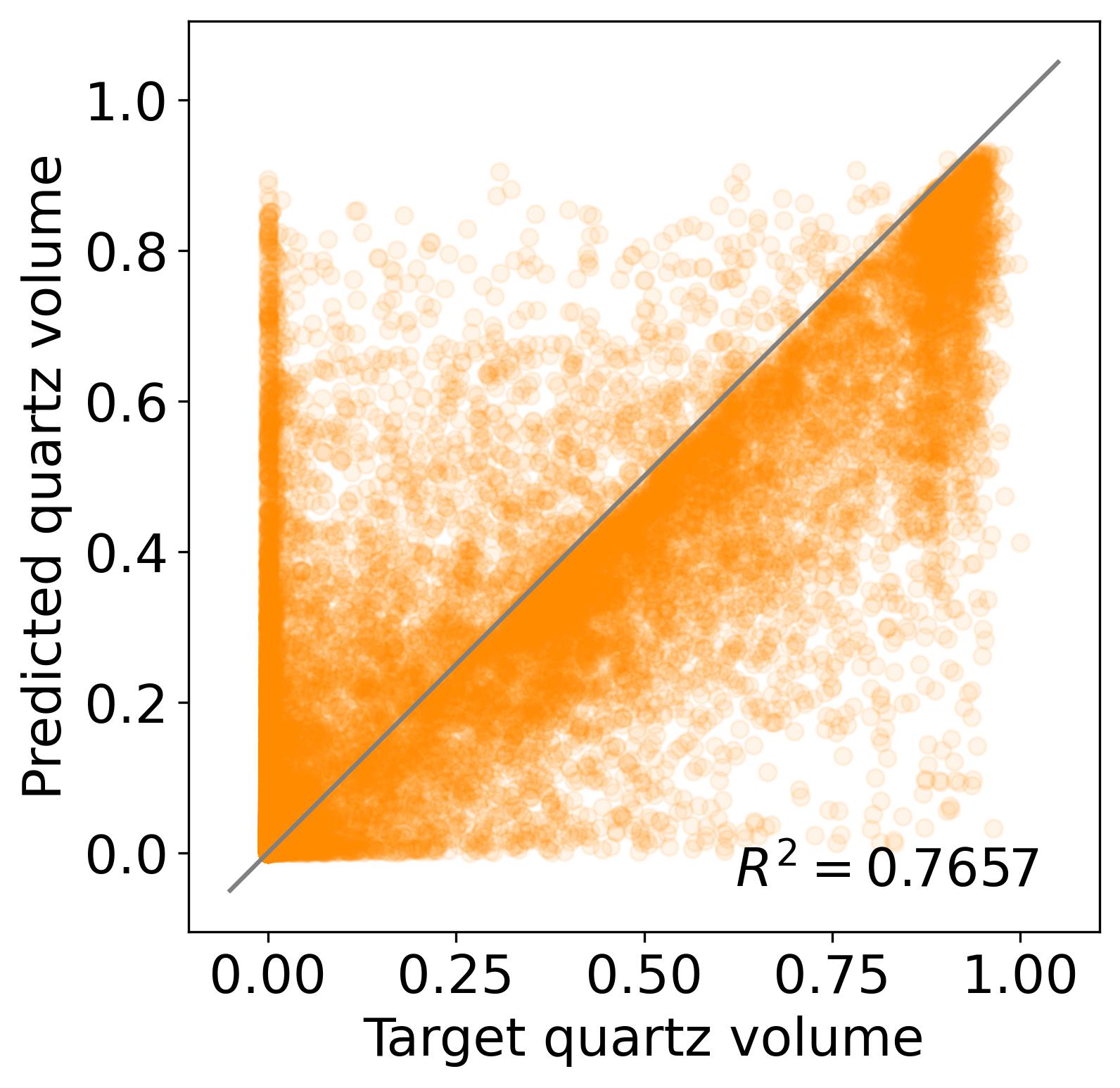} }}
    \subfloat[\textbf{(d)}\quad k = 1$\cdot10^{-12}$]{{\includegraphics[scale=0.3]{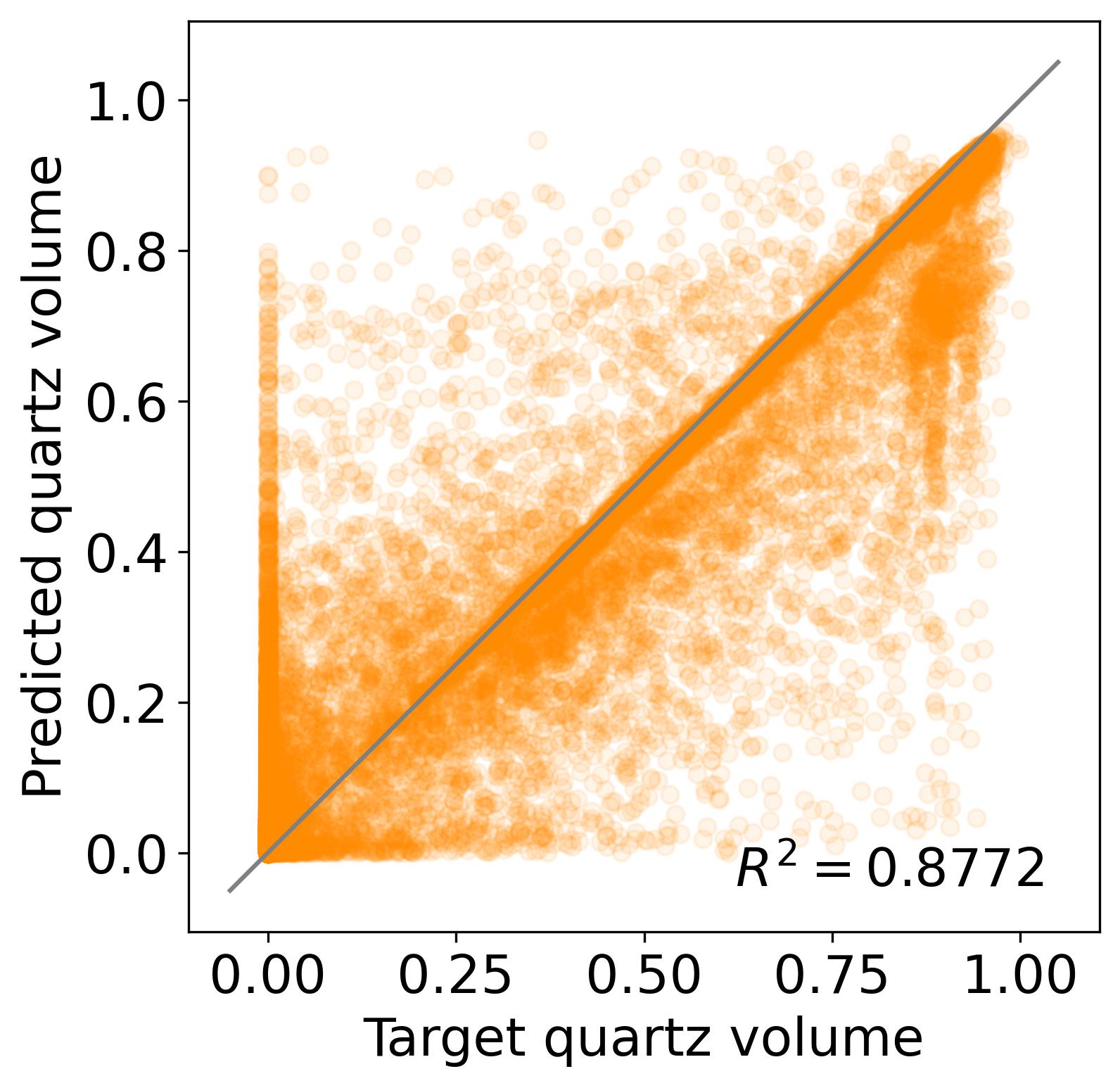} }}
    \caption{Random forest train and test predictions for each of the rate constants using all features (geometrical, topological, and hydrological). Each column depicts the results for a different rate constant. The regression model can predict the remaining quartz volume with higher accuracy for lower rate constants. The reaction rate constant is in [\textbf{mol m}$\mathbf{^{-2}}$ \textbf{s}$\mathbf{^{-1}}$] units and each train/test model has an equal number of data samples.}%
    \label{fig:rate_const_single}%
\end{figure}

In order to have a better grasp on why training a regression model using all rate constants does not deliver a very accurate prediction for the quartz volume left in each fracture, we train four more random forest regression models using all input features and one rate constant at a time. The coefficient of determination (R2 value) during training and testing is depicted in Figure~\ref{fig:rate_const_single}. 
One can see that the random forest regression model has difficulties predicting the remaining quartz volume for large rate constants, (k) of 1$\cdot10^{-9}$ and 1$\cdot10^{-10}$ mol m$^{-2}$ s$^{-1}$, while for small rate constants, (k) of 1$\cdot10^{-11}$ and 1$\cdot10^{-12}$ mol m$^{-2}$ s$^{-1}$, the model performs significantly better. When the rate constant is large, we expect more rapid quartz dissolution as fresh water is introduced to the system. This results in advection-dominated flow and quartz depletion in the backbone of the DFN, leading to shorter reactive transport simulation times since the quasi-steady state is reached in less time. It might seem a bit counter-intuitive, but for large rate constants the quartz volume in the system will be either flushed out completely from the fracture (mostly in a close proximity to the backbone, advection-dominated flow), or it will be clogging the fractures to a high extend (mostly in the secondary sub-network, diffusion-dominated flow), leading to sharp interfaces between the primary and the secondary sub-networks. This effect can be seen in Figure~\ref{fig:rate_const_single}(a) and (b), where most of the samples are located either in the lower left or in the upper right corners. The figures seem to have fewer data samples, but this is not the case, the data points are simply overlapping in the aforementioned regions. This makes predicting the remaining quartz volume for a single fracture much harder for the regression model since the distance to the backbone feature will carry less significant information. 

Reaching a quasi-steady state for simulations with lower reaction rates, (k) of 1$\cdot10^{-11}$ and 1$\cdot10^{-12}$ mol m$^{-2}$ s$^{-1}$, takes longer times since the reactions within fractures are slower. For those systems, we observe much smoother interfaces between the quartz-filled and quartz-depleted regions. This allows a smoother transition between these regions and more variability in the distance to the backbone feature, making the prediction of the remaining quartz volume an easier task. This is displayed in Figure~\ref{fig:rate_const_single}(c) and (d), where the data samples are almost equally distributed in the range of quartz volume. In this case, the regression model predicts the remaining quartz volume with higher accuracy.

\section{Discussion and Conclusions}
\label{sec:discussion}
We have provided a set of numerical simulations along with random forest regression models to characterize the interplay of the network geostructure and geochemical reactions in fractured media. We generated a set of generic three-dimensional fracture networks composed of a single families of mono-disperse disc-shaped fractures. We simulated flow and reactive transport using four reaction rate constants, (k) of 1$\cdot10^{-9}$, 1$\cdot10^{-10}$, 1$\cdot10^{-11}$ and 1$\cdot10^{-12}$ mol m$^{-2}$ s$^{-1}$, to determine the important features that control mineralization in the form of dissolution. Prior to the simulation, the fractures are filled with quartz that dissolves gradually during the simulations until a quasi-steady state is reached. The DFN topology and the reaction rate control the amounts of quartz that remain in each fracture, which strongly differs within a fracture network. In order to understand better what causes these discrepancies, we combine DFN graph representation and machine learning techniques, in particular random forest regression models. We constructed two types of regression models: (1) using all rate constants while varying the number of feature categories (topological, geometric, and hydrological); and (2) using all available features and training the model only on one reaction rate at a time. 

The first type of ML model assessed the input feature importance for each of the models. The results showed that building a regression model to predict the remaining quartz volume for all reaction rate constants is a difficult endeavor. We can conclude that including topological and geometric features is instrumental in building a useful regression model predicting the remaining quartz volume in a quasi-steady state. Including the hydrological features are the least important ones since they only slightly improve the accuracy of the regression models. The most important features controlling the quartz dissolution are the rate constant and the distance to the backbone (primary sub-network), which can be easily explained since they control the flow channelization and the strength of the chemical reaction, respectively.

The second type of ML model showed how the rate constant controls the dissolution in the DFN. The random forest regression model is able to predict the remaining quartz volume much better for low reaction rates in comparison to high reaction rates. This is due to smoother interfaces between the quartz-filled and quartz-depleted regions in the simulations using low rate constants, allowing for more variability in the quartz volume with respect to the distance to the backbone, which helps to build a better model.

In conclusion, we would like to summarize our findings:
\begin{itemize}
    \item The reaction rate constant and the distance to the backbone are the two most important features for predicting the remaining quartz volume in a single fracture. 
    \item The model has a hard time predicting the quartz volume remaining in a fracture using only topological features. However, adding geometric features improves the predictions significantly. Interestingly,  including the hydrological features as well only slightly improves the regression model predictions. This indicates that each of these features sets contributes to the predictions in meaningful ways. In all models the distance to the backbone (a topological feature) was the most useful for prediction. This feature thereby provides a foundation for the other properties to refine upon. 

    \item We performed an extensive cross-validated grid search. The optimized model delivers slightly better results compared to the base model. Even though the optimized model is much slower than the base model (approx. 22 times), it is more robust and prone to overfitting. 
    \item High reaction rates, (k) of 1$\cdot10^{-9}$ and 1$\cdot10^{-10}$ mol m$^{-2}$ s$^{-1}$, result in sharper interfaces between the quartz-filled and quartz-depleted fractures within a DFN while lower reaction rates, (k) of 1$\cdot10^{-11}$ and 1$\cdot10^{-12}$ mol m$^{-2}$ s$^{-1}$, lead to smoother interfaces of quartz volume between the quartz-filled and quartz-depleted fractures within a DFN. In the former the backbone is completely depleted, while the fractures in the secondary sub-network are still filled with quartz. These reactive transport simulations take a shorter time to reach a quasi-steady state. The distance to the backbone, which we observe to be one of the most important features predicting the remaining quartz volume, becomes less significant as the quartz volume associated with it is either 0 or close to 1, as there is almost no variability in the overall network. In the case of the lower reaction rates, the simulation takes longer times to reach a quasi-steady state and the freshwater can penetrate deeper into the secondary network and dissolve larger amounts of the quartz in these fractures. In these cases, the distance to the backbone and the quartz volume associated with a specific fracture carries more variability and as such serves as a stronger predictor of the quartz volume compared to the same feature when high reaction rates are used. The occurrence of sharp interfaces could become more important if precipitation is considered as the could lead to system clogging. 
\end{itemize}

As is slowly becoming better understood, our results reiterate that network connectivity is at the top of the hierarchy in determining flow and transport properties in fractured media. 
A primary contribution of this work is characterizing how that hierarchy influences reactions in fractured media as well.
To this end, we observe the dynamic reactions in the system exhibit a strong interplay with the topology, geometry, and hydrology of the network. 
In summation, the complex structure of the fracture network determines the flow field and subsequent, which is then dynamically modified by the reactions. 
In terms of implications, our work indicates that a critical step in improving predictive capabilities of reactive in fractured media remains improving our ability to characterize fracture networks. 
However, it is currently unknown how the inclusion of precipitation which could lead to blocking would influence these results. 
It is feasible that the feedback loop of precipitation leading to lowering permeability could alter the importance of our determined features. 
Thus, the next step in this line of research is two fold. 
First, we need to characterize how changing fracture network properties influences our predictive capabilities as well as feature importance. 
Second, a gradual increase in chemical complexity is required to take these insights into the field. 
Both of these avenues warrant further exploration and detailed research.


\section*{Acknowledgements}
JDH, MRS, and HSV thank the Department of Energy (DOE) Basic Energy Sciences program (LANLE3W1) for support. JDH, MRS, and HSV also gratefully acknowledge support from the LANL LDRD program office Grant Number \#20220019DR. AP acknowledges the Center for Non-Linear Studies at Los Alamos National Laboratory. Research was supported as part of the Center on Geo-process in Mineral Carbon Storage, an Energy Frontier Research Center funded by the U.S. Department of Energy (DOE), Office of Science, Basic Energy Sciences (BES), under Award \#DE-SC0023429. Los Alamos National Laboratory is operated by Triad National Security, LLC, for the National Nuclear Security Administration of U.S. Department of Energy (Contract No. 89233218CNA000001). 
Assigned LA-UR-23-33803.

\printcredits

\bibliographystyle{unsrtnat}

\bibliography{references}



\end{document}